\documentclass[11pt,aps]{revtex4-1}

\usepackage{graphicx}
\usepackage{wrapfig}
\usepackage{natbib}
\usepackage{xcolor}
\usepackage[normalem]{ulem}

\newcommand{\be}{\begin{equation}}
\newcommand{\ee}{\end{equation}}
\newcommand{\nn}{\mbox{} \nonumber \\ \mbox{} }
\newcommand{\ba}{\begin{eqnarray}}
\newcommand{\ea}{\end{eqnarray}}
\newcommand{\om}{\omega}

\newcommand{\E}{{\bf E}}

\renewcommand{\v}{{\bf v}}

\newcommand\eg{{\it{{e.g., }}}}

\newcommand{\Lf}{{Lorentz factor}}
\newcommand{\Lfs}{Lorentz factors}
\newcommand{\Bf}{{magnetic field}}

\newcommand{\Ef}{{electric  field}}
\newcommand{\Efs}{{electric fields}}

\newcommand{\NS}{neutron star}
\newcommand{\NSs}{{neutron stars}}
\newcommand{\EM}{electromagnetic}

\newcommand{\Schr}{Schr\"odinger}
\newcommand{\ms}{magnetosphere}
\newcommand{\mss}{magnetospheres}

\def\8{\infty}
\def\oh{\frac{1}{2}}

\def\d{\partial}

\def\undertext#1{\vtop{\hbox{#1}\kern 1pt \hrule}}

\def\VEV#1{\left\langle\,#1\,\right\rangle}

\def\dd#1{\frac{d}{d#1}}

\def\pp#1{\frac{\partial}{\partial#1}}
\def\pbyp#1#2{\frac{\partial#1}{\partial#2}}

\def\be{\begin{equation}}
\def\ee{\end{equation}}
\def\bea{\begin{eqnarray} & &}
\def\eea{\end{eqnarray}}

\def\rf#1{(\ref{#1})}

\begin{document}

\title{Anderson self-localization  of light in pair plasmas}

\author{Maxim Lyutikov $^1$, Victor Gurarie  $^2$ \\
 $^1$ Department of Physics and Astronomy, Purdue University, \\  525 Northwestern Avenue, West Lafayette, IN 47907-2036, USA\\
 $^2$ Department of Physics and Center for Theory of Quantum Matter, \\ University of Colorado, Boulder CO 80309, USA }

\begin{abstract}
We demonstrate that in pair plasma weakly nonlinear electromagnetic waves, $a_0 \leq 1$,   experience Anderson self-localization. The beat between the driver and a back-scattered wave creates  charge-neutral,    large random,  {yet correlated} density fluctuations  $\delta n/n_0  \gg 1$, and  corresponding  fluctuations of the dielectric permittivity $\epsilon$   (random plasma density grating).  Propagating in quasi-1D, waves in a medium with spatially random  self-created fluctuations of dielectric permeability experience localization. In the linear regime,  the instability  can be classified  as Induced Brillouin Scattering; it is  described by the parameter   $\rho _L  =  \left(    a_0 {  \om_{p}/ }{\om}\right)^{2/3} \ll 1 $,
related to  the Pierce parameter  of Free Electron Lasers.
In the cold case, {the growth rate is $\Gamma \approx  \rho _{L}  \om \ll 1 $}  ($a_0 $ is laser nonlinearity parameter, $\om_p$ is plasma frequency,  $\om$ is the laser frequency).  
Anderson self-localization of light leads to (i) reflection of EM waves by the under-dense pair  plasma; (ii)  a wave already present inside the  plasma separates into bright trapped pockets and dark regions.  Mild initial  thermal spread  with $\Theta \equiv k_B T/(m_e c^2) \approx a_0^2$, restores  wave propagation by suppressing the seeds of parametrically unstable density fluctuations.  A circularly polarized driver produces   linearly polarized  structures, with position angle varying randomly between the bright  pulses. {Time-variability of the resulting density structures does not suppress localization due to remaining correlations (not white noise)}. We discuss possible applications to astrophysical Fast Radio Bursts. 
\end{abstract}

\maketitle

\section{Introduction}

Anderson localization \citep{1958PhRv..109.1492A} is a broad term used to describe  phenomena where waves become
trapped when propagating in disordered media. Unlike in a classical diffusion process, the interference of scattered waves is important for waves to become Anderson-localized.

Here we are concerned with a related classical (non-quantum) problem of localization of light in random media
\cite{doi:10.1137/1104038,1991PhT....44e..32J}. We address a new, unusual regime of {\it self}-localization, when randomness of the  dielectric properties of the medium - pair plasma in the particular case -  is created by the wave itself.

Motivation for this work comes from the importance of ultra-strong laser-matter interaction, both in the laboratory and in the field of relativistic plasma astrophysics. Modern lasers with fluctuating \Ef ~$E_w$ 
can achieve nonlinear intensities
\be
a_ 0  = \frac{e E_w}{m_e c  \om} 
\ee
in the hundreds. (Parameter $a_0$ is approximately the dimensionless momentum of jitter motion of a particle in an EM wave.) High-intensity colliding laser pulses can also lead to the creation of abundant electron-positron pairs \citep{2006RvMP...78..591M,2008PhRvL.101t0403B,2020PhPl...27e0601Z}.

In addition to being of fundamental interest, the physics of ultra-strong laser-matter interaction became a forefront research topic in relativistic plasma astrophysics, initiated by meteoric developments over the last years in the field of mysterious Fast Radio Bursts (FRBs) \citep{2022A&ARv..30....2P}. The plasma physics challengers are enormous: How to produce and propagate ultra-intense millisecond-long radio bursts from $\sim $  halfway across the visible Universe \citep{2007Sci...318..777L,2016MNRAS.462..941L,2016Natur.531..202S,2022A&ARv..30....2P,2019ARA&A..57..417C}. 

 An important astrophysical hint comes from the
observations of correlated radio and X-ray bursts from ultra-magnetized \NSs\ \citep{2020Natur.587...54C,2021NatAs...5..372R,2020Natur.587...59B,2020ApJ...898L..29M,2021NatAs.tmp...54L}. This 
established the  FRB-magnetar connection. If the emission originates in the \mss,  the laser non-linearity parameter in these settings can be as high as staggering  $a_0 \sim 10^9$. 
Generation of FRBs, \eg\ within {\NSs}' magnetospheres, 
as well as the escape of high-intensity radiation 
\cite{2022PhRvL.128y5003B,2024MNRAS.529.2180L} remain topics of active research. 

There is a great deal of analogy between quantum and EM systems, as both are described in terms of wave phenomena. 
A principal difference between typical quantum mechanical and classical EM systems with disorder lies in the fact that  the \Schr\ equation is first order in time, while the wave equation is of the second order in time. Even in the stationary case,  when we can Fourier transform in time, the 1D \Schr\ equation with potential $V(x)$ and wave equation with varying dielectric constant $\epsilon  (x)$ look qualitatively distinct:  in the case of the classical wave equation, the perturbing function is multiplied by $\om$. This implies that the strength of localization is smaller at low frequencies. On the other hand, in a random dielectric, the phase coherence needed for Anderson localization is harder to achieve at higher frequencies/shorter wavelengths.   Though in a random 1D medium localization always occurs, in an astrophysical dynamical setting,  the interaction time is limited.
Thus, we expect that localization occurs in a limited range of frequencies -  not too high and not too low.

In the present work, we address the most basic problem: propagation of strong \EM\ waves in pair plasma. Although the topic has a long history \cite{1974PhRvL..33..209M,1974PhRvL..33..209M,1975OISNP...1.....A,1989JPlPh..42..507K,2016PhRvL.116a5004E}, as we argue below, important points were partially overlooked. 

{
As the work combines interconnected effects  from various sub-fields of physics, let us remind the reader of the basic definitions. {\it Parametric instability} here refers to a nonlinear process whereby the wave amplifies itself the small perturbations of its amplitude. 
{\it  Induced scattering} (or nonlinear scattering) refers to the process that involves many particles and many waves (in a phase correlated way). In pair plasma, since the effects of charge separation are small, the dominant process can be classified as induced Brillouin Scattering - scattering on overall charge-neutral density fluctuations (but not necessarily propagating sound or ion-acoustic  waves). The dominant growing mode has a wavelength half  that of the driver. In a tenuous medium,  waves back-scattered by obstacles separated by half wavelength  add coherently and thus amplify the reflected signal -  this effect is known as coherent Bragg's back-scattering.  At the nonlinear stage we consider, there is no clear separation between these different processes.
}


{
The plan of the paper is as follows.  In \S \ref{Reflection} we discuss the key result: nearly complete reflection of a weakly nonlinear EM wave by under-dense  pair plasma (PIC simulations).   In \S \ref{InducedBrillouin} we consider the linear stage of the instability and defined an important parameter $\rho_L$, Eq. (\ref{rhoL}). 
In \S \ref{deltaspike} we consider a complementary problem: scattering of EM waves by a collection of static density  $\delta$-spikes: this illustrates
Anderson localization of EM waves in pair plasma with random density sheets.
In \S \ref{inside} we consider
coherent back-scattering and localization of waves present and/or growing  inside the medium. In \S  \ref{Clemmow1} we  use PIC simulations to consider localization  of  waves already inside plasma 
}

\section{Reflection of weakly nonlinear EM wave by under-dense pair plasma: PIC simulations}
\label{Reflection}

In this section, we investigate the dependence of the localization parameters on the waves' nonlinearity $a_0$. We concentrate on mild non-linearity parameter $a_0 \leq 1$; see Ref. \cite{2025arXiv250906230T} for the case of $a_0 \geq 1$. At mild  $a_0\leq 1$ there are  two competing effects. First,
smaller $a_0$ reduce the strength of interaction. Second,  larger $a_0$ lead to bulk acceleration of pair plasma and corresponding density pile-up; these effects   complicate interpretation of the results. 

In this Section, we start with progressively higher $a_0= 10^{-2},  10^{-1} $. { 
We draw two  important conclusions: (i) long  \EM\ pulses, extending beyond (\ref{L}),   are reflected by pair plasma; (ii) as a result of the reflection, 
 in pair plasma bulk acceleration by the EM pulse  switches to the snowplow regimes, as opposed to ponderomotive push,  as early as $a_0=10^{-1}$,  \S \ref{bulk}. 
At smaller intensities, bulk acceleration is intermediate between the two regimes.
}

\subsection{Weakly  non-linear $a_0 =10^{-2}$ EM pulse}

We start with one of the most striking of our results: reflection of a weakly nonlinear EM wave by an under-dense pair plasma. Unlike previous work \citep{2025arXiv250906230T}, we limit the calculations to 
   relatively small  intensities,  $a_0 =10^{-2}-10^{-1}$, a wave which is only weakly non-linear. The reason for this particular choice of $a_0$  is important and subtle, { see \S \ref{bulk} and Eq. (\ref{a0c}).} 

\begin{figure}[h!]
  \centering
 \includegraphics[width=.99\linewidth]{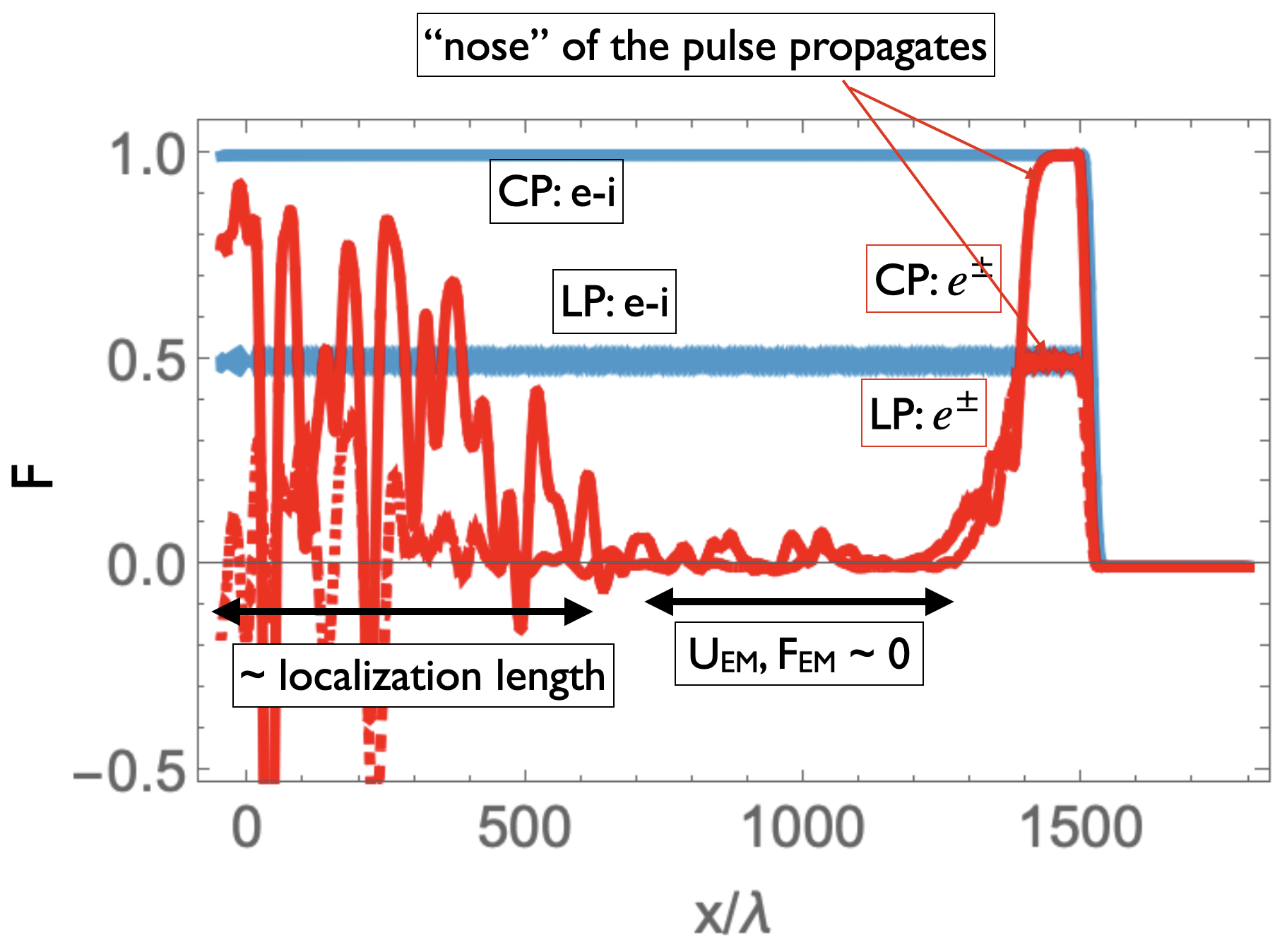}
 \includegraphics[width=.49\linewidth]{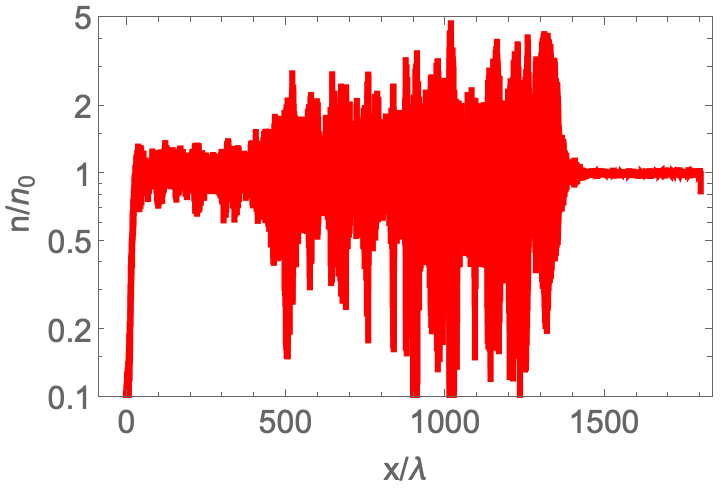}
  \includegraphics[width=.49\linewidth]{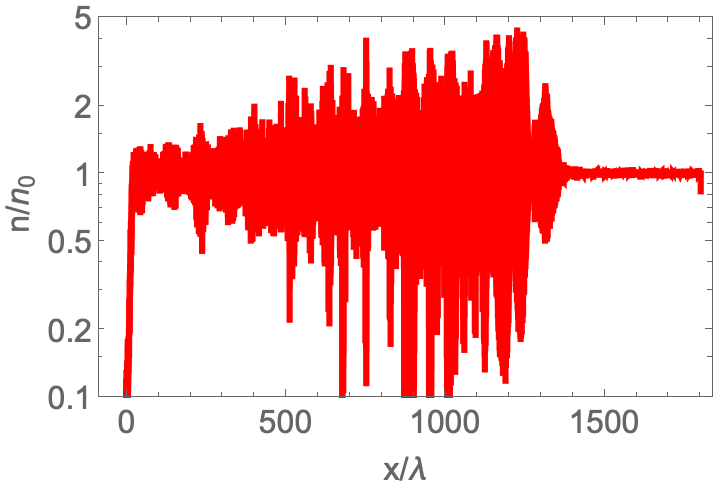}
\caption{Interaction of weakly nonlinear waves $a_0 =10^{-2}$ with  under-dense $n = 0.0125 n_{\rm cr}$ plasmas. Top panel: values of Poynting flux $F$ normalized to the incoming one for CP (so, LP is two times smaller);  red lines are for pair plasma,  blue lines for e-i plasma;  circular (CP) and linear (LP) polarizations.  Bottom row: density for CP (left) and LP (right) for the $e^\pm$ plasma. We observe  huge density variations $\delta n/n_0 \gg 1$. }
\label{a001}
\end{figure}

We find that already at $ a_0 \sim  10^{-2}$,  the effects of nonlinearity are  hugely important for pair plasma. In Fig. \ref{a001}, we show the results of PIC simulations of weakly nonlinear waves, $a_0 =10^{-2}$, falling onto electron-positron and electron-ion plasma. 
For simulation details and convergence test, see Appendix \ref{Epoch};  also see  Fig. \ref{PIC-plus-minus}.

{
For the e-i case,  waves propagate within the plasma as expected (blue lines in Fig. \ref{a001}. In contrast, in the $e^\pm$ plasma, after the initial ``nose'' of the wave propagates into the plasma, the bulk of the wave is reflected. As we discuss in the present paper, this effect is due to the effect of  Anderson self-localization of light in a pair plasma.

Nonlinear effects (\eg density fluctuations in a LP wave) are expected to scale as $\propto a_0^2$. For our parameters, it is then expected that $\delta n/n_0  \approx 10^{-4}$. Indeed, in the electron-ion case (blue lines in Fig. \ref{a001}) density fluctuations are small. The pair plasma cases is very different: there are large, $\delta n/n \sim 10$ density fluctuations (bottom row). 
}

 \begin{figure}[h!]
  \centering
 \includegraphics[width=.99\linewidth]{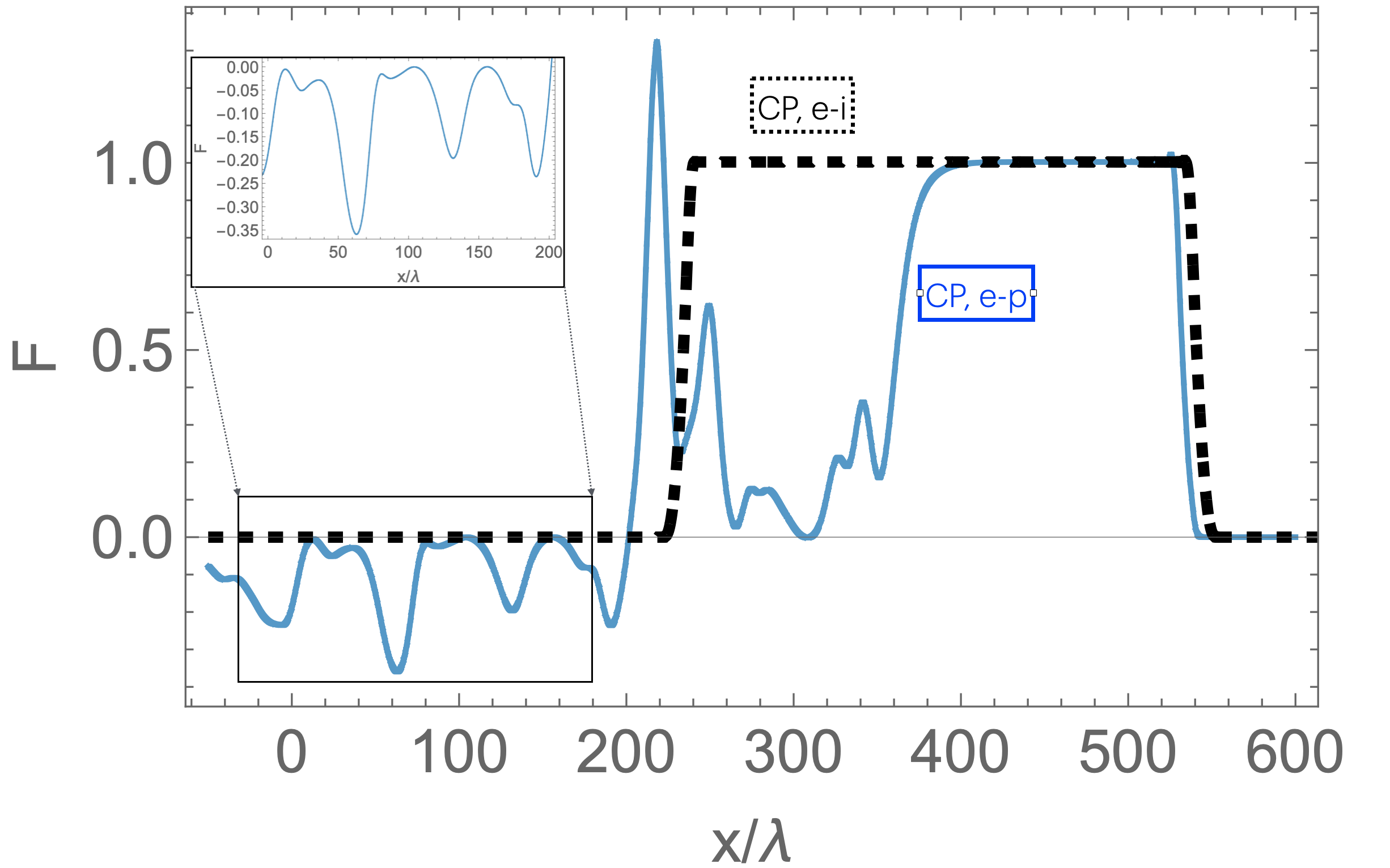}
\caption{Reflection of a CP EM pulse by under-dense pair plasma (solid blue curve) if compared with e-i case (black dashed). Negative values of Poynting flux (zoomed insert) indicate reflection. Same parameters as Fig. \ref{a001}. {In the pair plasma case}, persistently negative Poynting flux is observed when the head to the pulse propagated thousands of wavelengths into the plasma.}
\label{Anderson-localization-PIC003}
\end{figure}

{\bf
In Fig. \ref{Anderson-localization-PIC003} we show results of PIC simulations of a circularly polarized  EM {\it  pulse}  (with a trapeze form, rump-up and rump-down) for two cases: electron-ion  (dashed line) and $e^\pm$ plasm (solid lines).
}

Fig. \ref{Anderson-localization-PIC003} shows that the e-i case (dashed line) behaves as expected: the pulse propagates. The $e^\pm$ case is very different: while the ``nose'' part propagates, most of the pulse is reflected.
A zoomed-in inset highlights the fact that there is a negative energy flux behind the pulse, indicating reflection. 
Negative Poyting flux drains energy from the bulk of the pulse, but not from the penetrating ``nose''. 

\begin{figure}[h!]
  \centering
 \includegraphics[width=.99\linewidth]{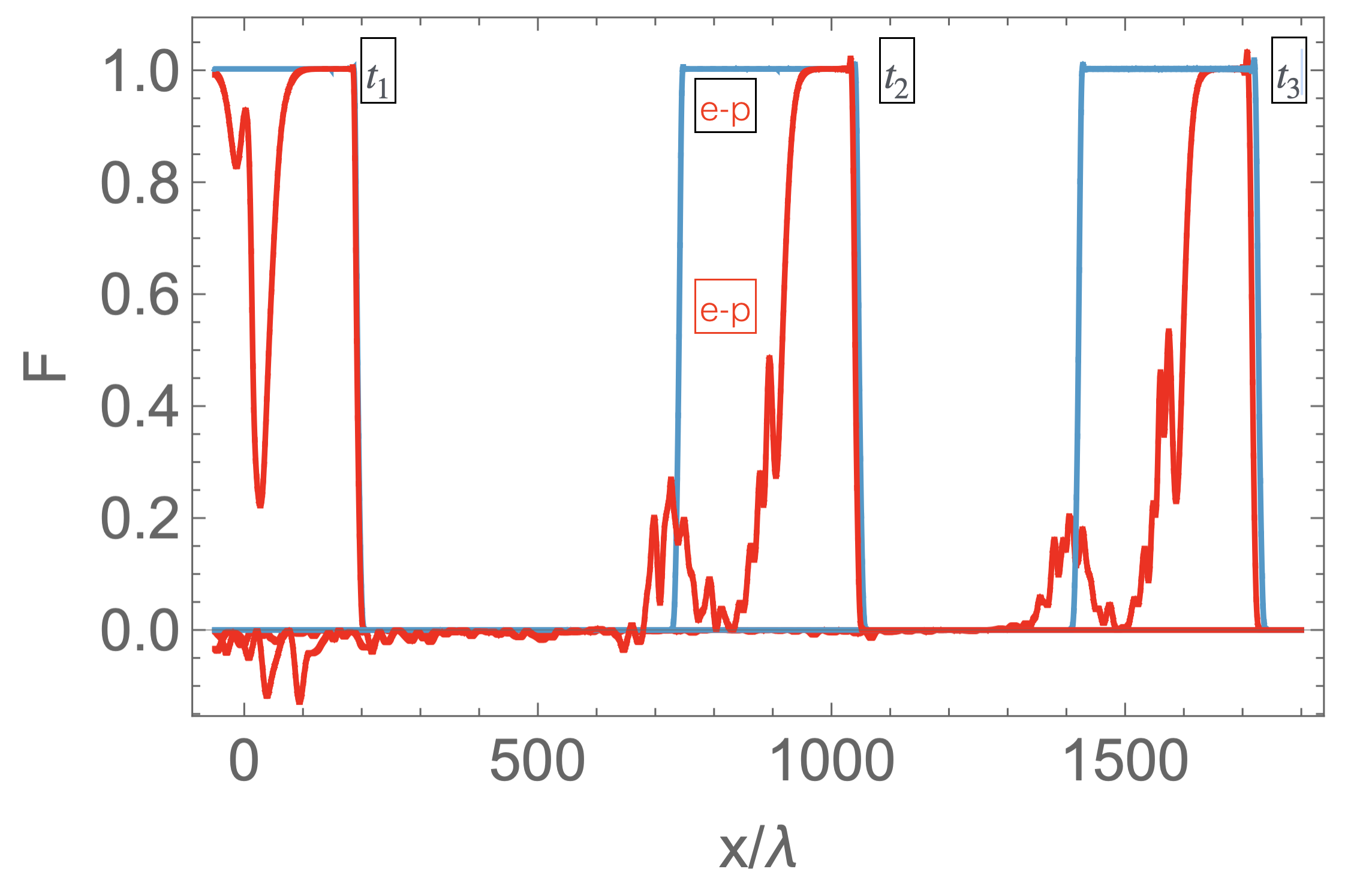}
\caption{Temporal evolution of a pulse in e-p and e-i cases; CP. $a_0=10^{-2} $. Shown are Poynting fluxes at different times $t_1, \, t_2, \, t_3$; blue line: e-i case, red lines: e-p case.}
\label{time}
\end{figure}

In Fig. \ref{time} we compare temporal evolution of a pulse in e-p and e-i cases.
Draining/reflection  of the pulse occurs on scales much larger than the wavelength, starting at hundreds and  continuing into thousands in the particular case. Such effects are typically termed mesoscale phenomena. { (For analytical estimate of the penetrating length, see Eq. (\ref{L}).)} 

\begin{figure}[h!]
  \centering
 \includegraphics[width=.99\linewidth]{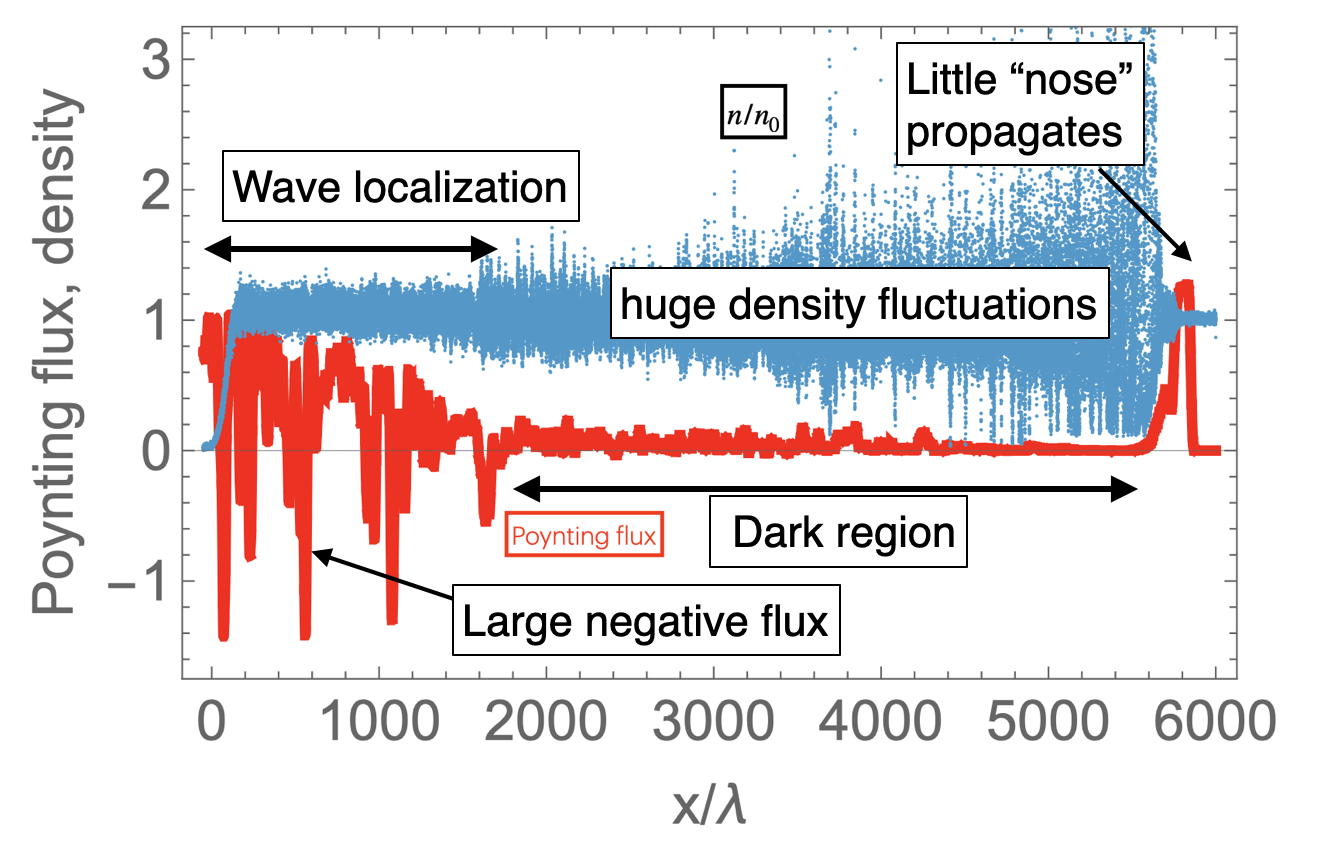}
\caption{Annotated longer simulations of CP pulse with $a_0 =10^{-2}$ propagating in pair plasma. Shown are Poynting flux (red curves) and density (dark blue), both normalized to the incoming flux and unperturbed density.  The EM wave penetrates into plasma only for a finite distance; the wave is Anderson-localized. For this run $n_x =20$. {\bf Though density fluctuations look random - they are not, see Fig \ref{dens0019_fft_zoom}.}}
\label{longCP-ep-001}
\end{figure}

 In Fig. \ref{longCP-ep-001} we show results of longer simulations of a CP signal propagating into a pair plasma. One clearly identifies the effects of Anderson localization: { the flux decays on scales $\sim $ few hundred wavelengths,} and after approximately $x/\lambda \geq 1000$ the Poynting flux is approximately zero. At smaller $x$ there are regions of large negative Poynting flux. Large fluctuations in random density are clearly seen. 

\subsection{Mild intensities $a_0 =0.1$: onset of Stimulated Raman Scattering  in e-i case, strong   bulk compression in $e^\pm$  case}
\label{001}

\begin{figure}[h!]
  \centering
 \includegraphics[width=.99\linewidth]{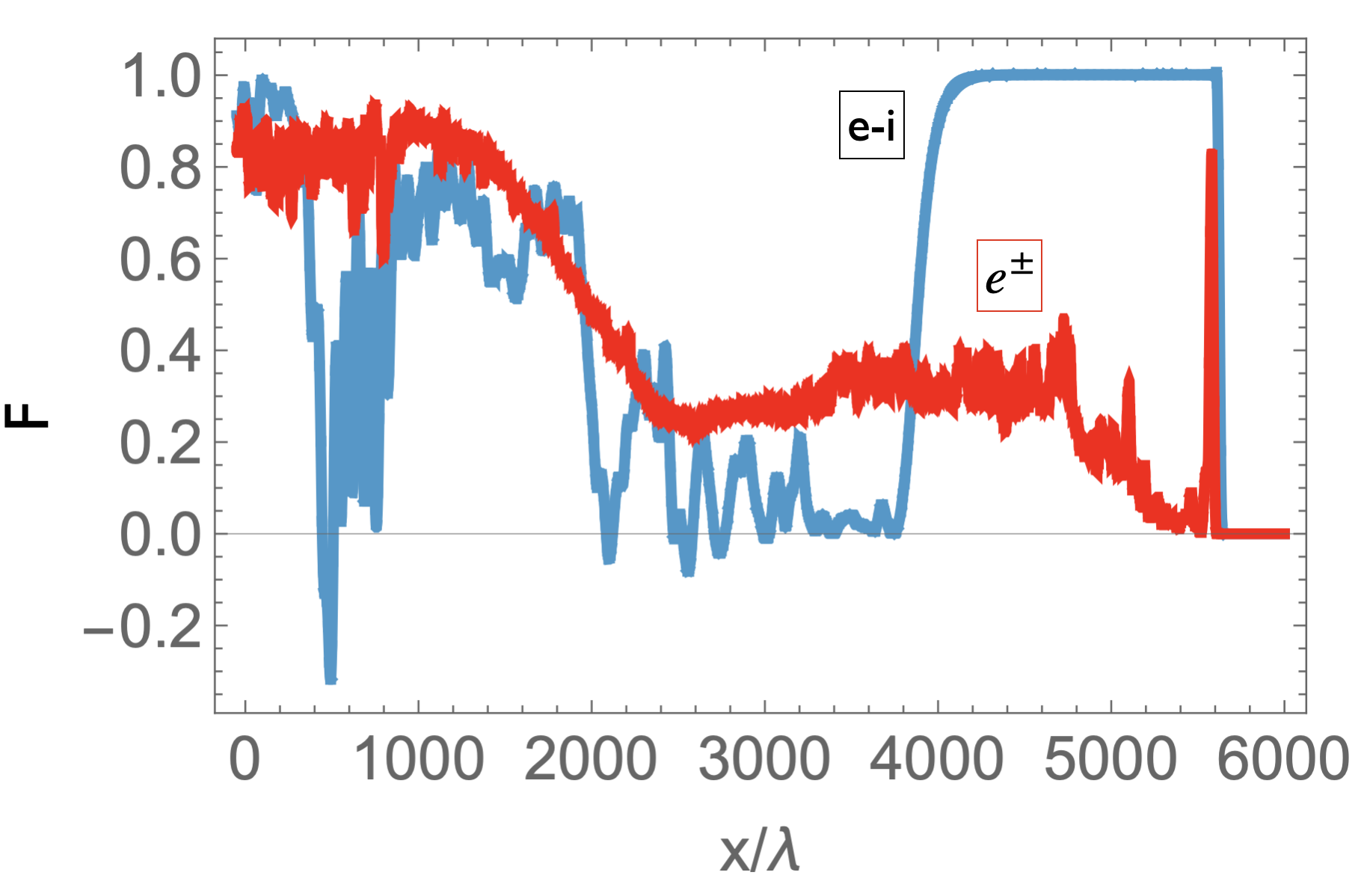}\\
  \includegraphics[width=.49\linewidth]{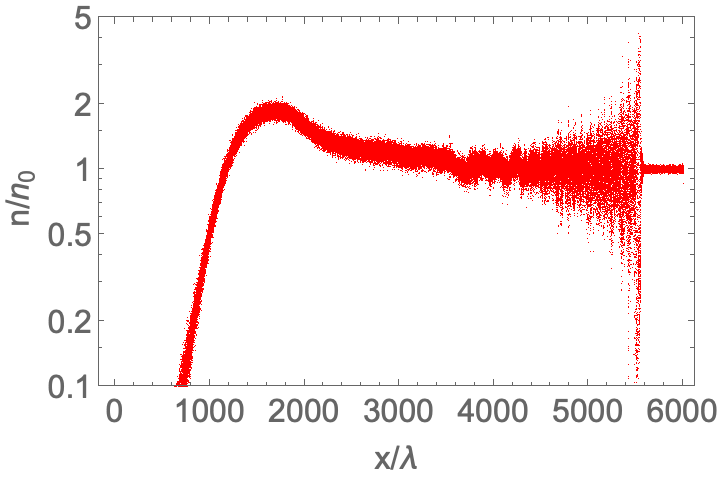}
   \includegraphics[width=.49\linewidth]{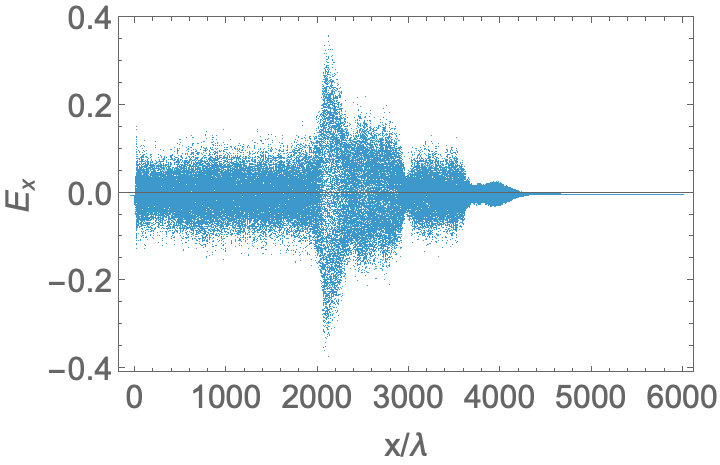}
\caption{The case of $a_0 =10^{-1}$, CP. {Top panel: Poynting  $F$ for $e-i$ and $e^\pm$ cases.} Bottom panel: density sweep-up for pair plasma (notice that even for $a_0^2  = 10^{-1}  \ll 1$ the resulting density increases is of the order of unity); parallel \Ef\ for e-i plasma (notice onset of plasma oscillations - SRS - corresponding to the dip in EM intensity in the range $ 2000 \leq x/\lambda  \leq 4000$). }
\label{a01}
\end{figure}

At relatively small, mild intensities,  $ a_0  =  10^{-1}$, 
Stimulated Raman Scattering (SRS) becomes important in e-i plasma, while in pair plasma the sweep-up becomes  very important, Fig. \ref{a01}. 

The $e^\pm$ case shows {\it large}, of the order of unity density increase due to plasma sweep-up, bottom left panel in Fig. \ref{a01}. (Charge density fluctuations in the $e^\pm$ case are relatively small, of the order of the numerical noise.)

The e-i regime shows onset of Stimulated Raman Scattering (SRS), around $x/\lambda \leq 4000$. This is accompanied by the conversion of the EM wave into plasma oscillations, bottom right panel in Fig. \ref{a01}.

\subsection{Bulk acceleration of pair plasma by in-falling EM wave}
\label{bulk} 

Our simulations indicate that pair plasma is strongly accelerated/swept-up even by mild laser pulse, $a_ 0 \leq 1$. Next we discuss different regimes of plasma sweep-up.

 For an EM wave falling onto plasma, one can qualitatively identify two regimes of bulk acceleration:  ponderomotive effects in the transparent case,   and momentum conserving/snow-plow in the reflection case. Ponderomotive acceleration (in the single-particle approximation)  follows from the conservation of generalized vector potential
 \ba &&
 \gamma = 1 + \frac{a_0^2 }{2}
 \nn && 
 \gamma_x = \frac{1}{\sqrt{1-v_x^2}} =\frac{2+a_0^2}{2 \sqrt{1+a_0^2}}
 \nn &&
 \beta_x= \frac{a_0^2}{2+a_0^2}
 \nn &&
 (  \delta n )/n_0 \approx a_0^2
 \label{pondero}
 \ea
 where $(  \delta n )/n_0$ is the expected density sweep-up.   Thus for $a_0 \leq 1$ one expects $\delta n/n_0 \approx  a_0^2 \ll 1$. For example, for $a_0 =10^{-1}$ the expected pile-up is $\sim 10^{-2}$. This  expectation is clearly contradicted by results of PIC simulations, Fig. \ref{a01} {bottom right panel}, where we observe $(  \delta n )/n_0 \approx 1$. 

 The reason for much larger density increase is that as EM waves are reflected, they transfer their momentum to the plasma. Such regime  of momentum conservation is sometimes called the ``snowplow" regime. 
 In the snowplow regime, when the wave is fully reflected, the bulk acceleration  is then well determined by momentum conservation \citep{2021PPCF...63d5010H}
\ba &&
\beta_x =  \frac{\sqrt{H} }{1+ \sqrt{H}}
\nn &&
H = 
\frac{a_0^2 \omega ^2}{2 \omega _p^2}= \frac{n_{cr}}{n_0} \frac{a_0^2}{2}
\label{snowplow}
\ea
where $I_0$ is intensity of laser pulse and $n_{cr}$ is the critical density corresponding to laser frequency.

The actual parallel velocity and density  increase  are  bracketed by 
(\ref{pondero}) from below and (\ref{snowplow}) from above.  
{Equating (\ref{pondero})  and  (\ref{snowplow}), 
the ponderomotive acceleration switches to ``snowplow" at
\be
a_0 \approx \frac{\om_p}{\om} \approx \sqrt{\frac{n_0}{n_{cr}}} \approx 0.1
\label{a0c}
\ee
where the estimate is for our typical run parameters.}

\section{Large density modulations created by EM wave in pair plasma}
\label{InducedBrillouin}

\subsection{Induced Brillouin  and Bragg's scattering  in pair plasma}

In what follows, we offer an explanation to the most unexpected of our results, the reflection of EM waves by an under-dense pair plasma.
The reflection of an EM wave by a pair plasma is due to  
the formation of {\it large} density spikes by the beat 
 between the driver and a back-scattered wave. Large density spikes and corresponding large variations of the dielectric constant, $\delta \epsilon \sim 1$, appear even for small nonlinearity, $a_0 \leq 1$.  Large, self-created by the wave itself, fluctuations of the dielectric constant then lead to the effect of Anderson localization.


Importantly, and very unusually for basic models of   Anderson localization, {\it density spikes are generated by the wave itself}, while usually they are statically imposed by ``an experimentalist" or by interactions with some additional time-independent random degrees of freedom, which are usually termed ``impurities").  The self-generation of density/ $\epsilon $ spikes leads to highly nonlinear, and importantly, time-dependent scattering/localization problem. {This can be called a regime of Anderson self-localization}.

\begin{figure}[h!]
 \includegraphics[width=.99\linewidth]{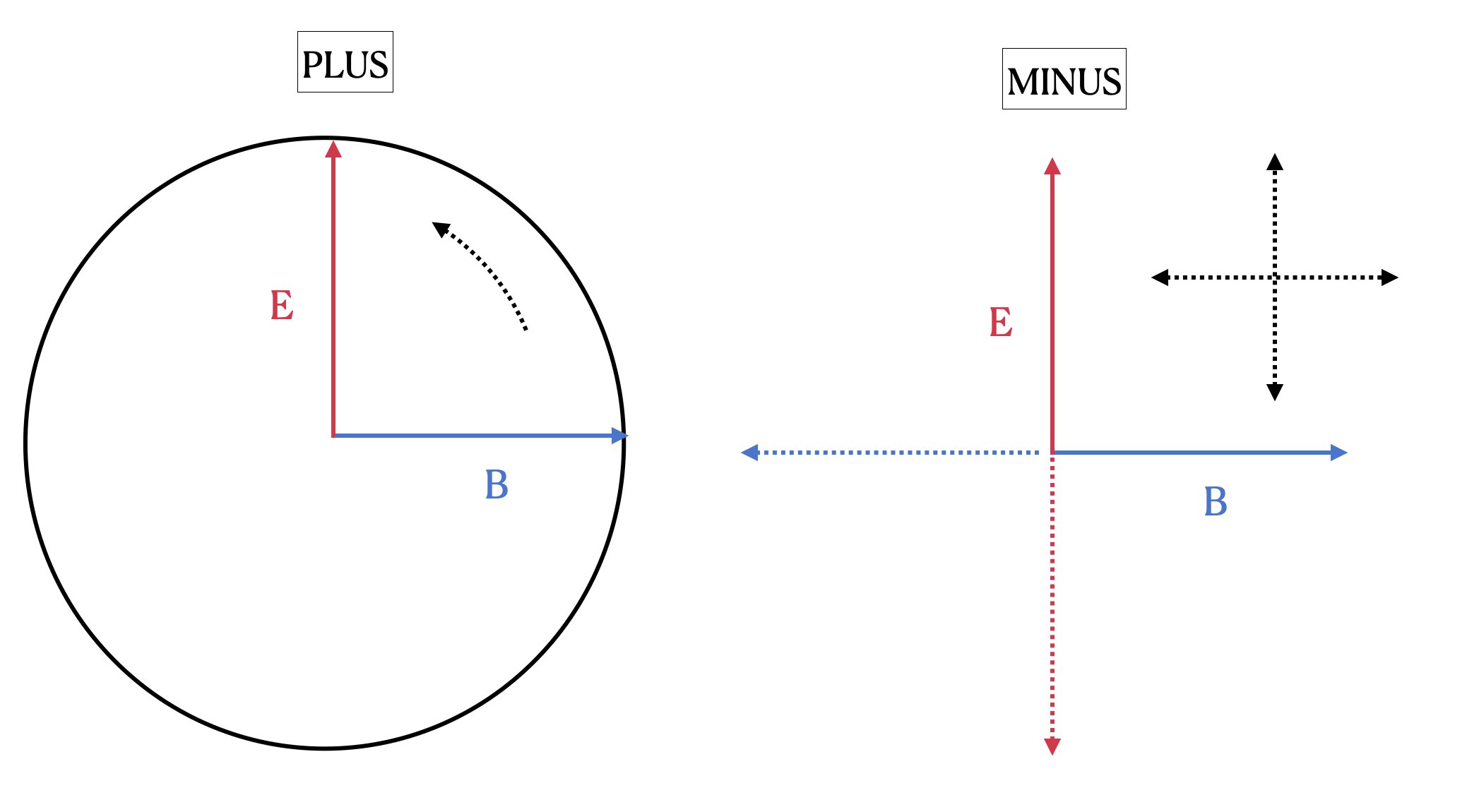}
\caption{Beat of electric and magnetic fields at a given location for  PLUS and MINUS   polarizations of the scattered wave for the exemplary case of equal amplitudes. In the PLUS configuration,  the \Ef\ and \Bf\ of the combined waves  rotate, while in the MINUS configuration, the  \Ef\ and \Bf\   oscillate (dashed lines). As a result, in the MINUS configuration the axial Lorentz force is spatially homogeneous, and no density bunches are created.
} 
\label{plus-minus}
\end{figure}

Formation of large density spikes can be understood with a simple example of particle motion in two counter-propagating waves, the driver and back-scattered wave, non-interacting particles, Ref. \cite{2025arXiv250906230T}. {Later, in \S \ref{linear} we consider analytically the linear stage of the correspoding parametric instability in two-fluid approach.}
Consider  a CP driver with  nonlinearity parameter $a_0$ and 
 a back-scattered wave with a nonlinearity parameter $a_1$.   There are two possible polarizations of the counter-propagating wave: (i) PLUS wave,   when both \Efs\ rotate in the same direction, as measured in absolute space (not with respect to their wave vector);
(ii)  MINUS wave, when \Efs\ rotate in the opposite direction, Fig. \ref{plus-minus}, see Ref. \cite{2021ApJ...922..166L}. 
Thus, PLUS combination produces circularly polarized standing  beat waves, while MINUS combination produces linearly polarized waves.

{
The two polarizations of the back-scattered wave,  PLUS and MINUS, have drastically different effects on the particle dynamics.
In the limit of weak perturbation, $a_1 \to 0$, the  slow axial dynamics of particles obeys
\ba &
\partial_t \delta v_x \approx  a_0 a_1 \om \sin ( 2 k x )  & \mbox{ PLUS}
\label{calA}
\\ &
\partial_t \delta v_x \approx a_0 a_1 \om \sin ( 2 t \om)  & \mbox{ MINUS}
\ea 
where $\delta v_x$ is the axial fluctuating velocity.
Thus, only the PLUS wave leads to the formation of bunches, on double wave number $2 k$. 
}

Most importantly, the jitter velocity (\ref{calA}) is independent of the sign of the charge (as a product of $a_0$ and $a_1$). In pair plasma, both charges oscillate in phase, without charge separation.
Typical frequency and amplitude of oscillations are
\ba && 
\Omega_p =  \sqrt{ 2 a_0  a_1} \om
\nn &&
\delta v_x/c=  \sqrt{ 2 a_0  a_1} 
\ea

\begin{figure}[h!]
  \centering
 \includegraphics[width=.3\linewidth]{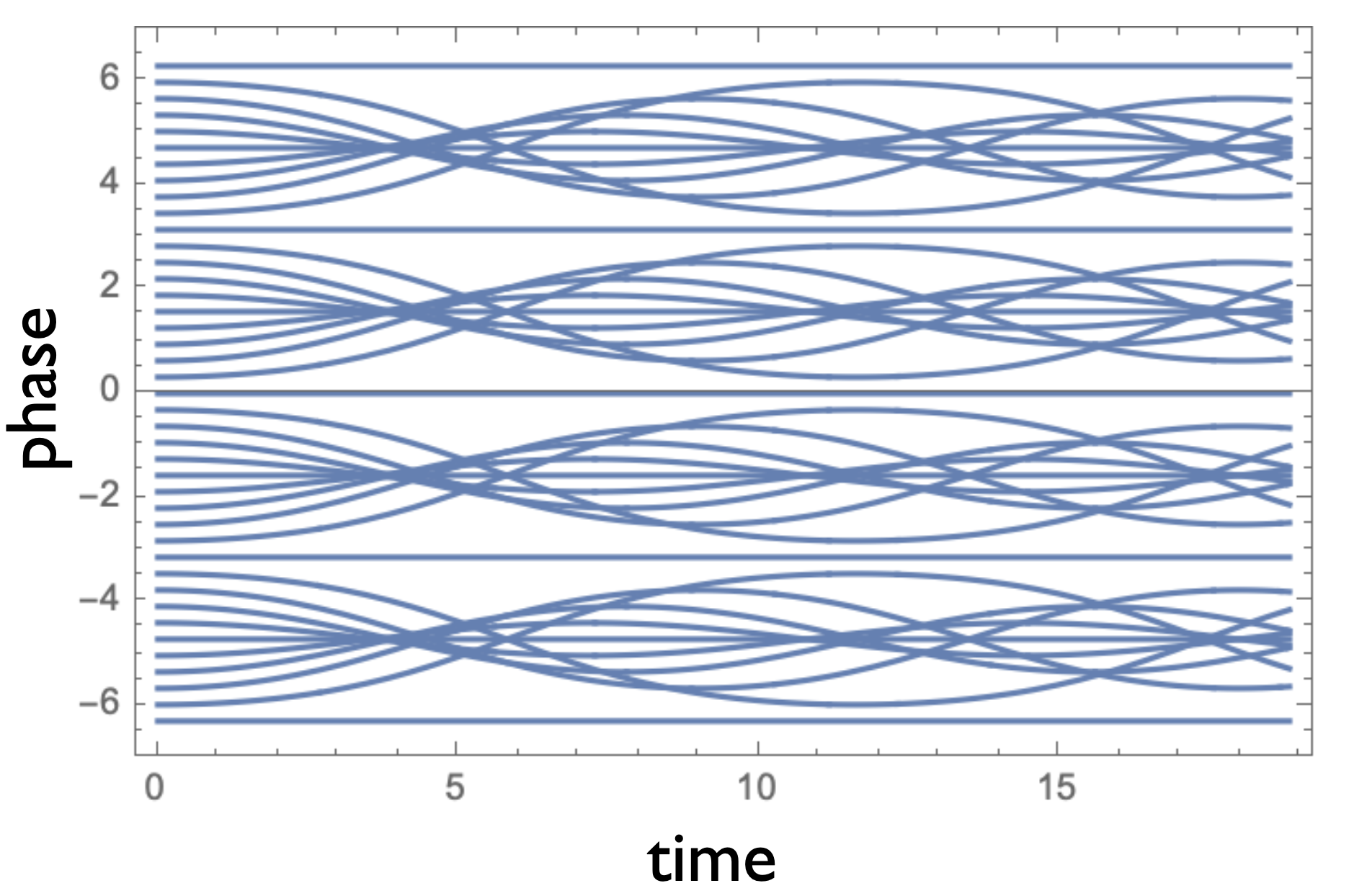}
 \includegraphics[width=.3\linewidth]{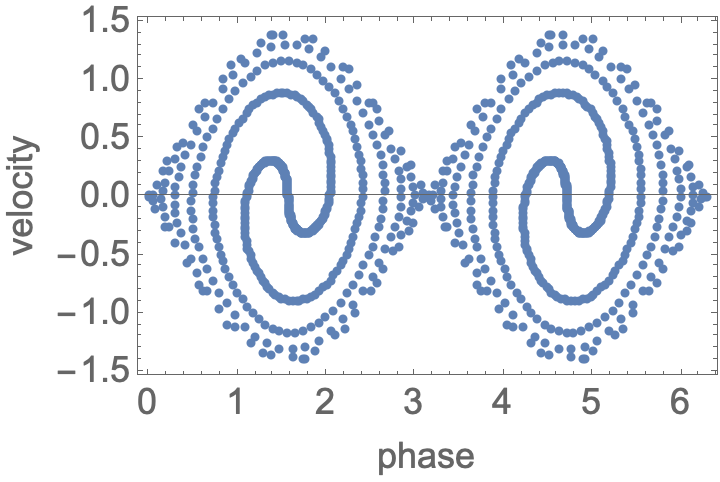}
  \includegraphics[width=.38\linewidth]{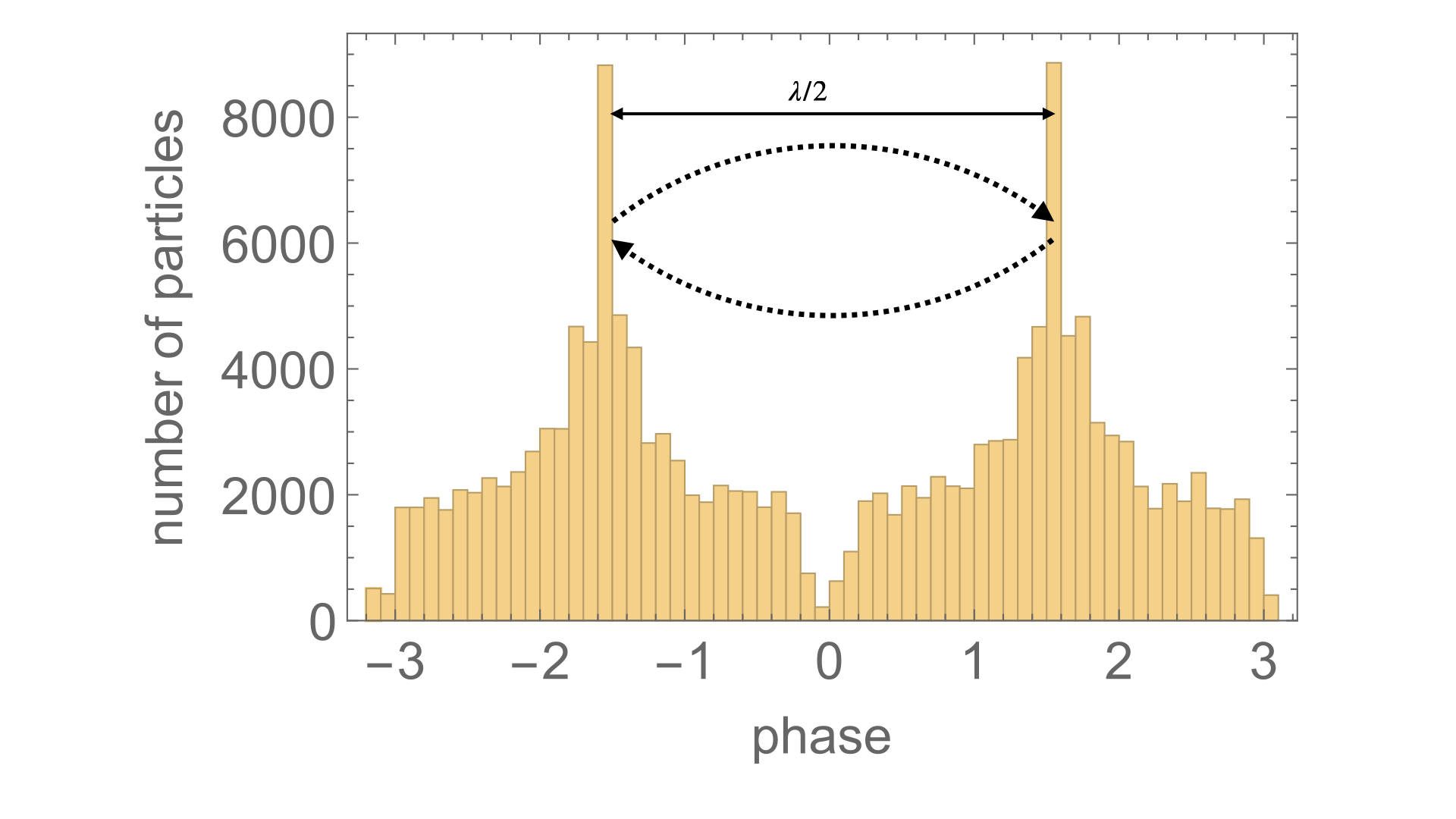}
   \includegraphics[width=.3\linewidth]{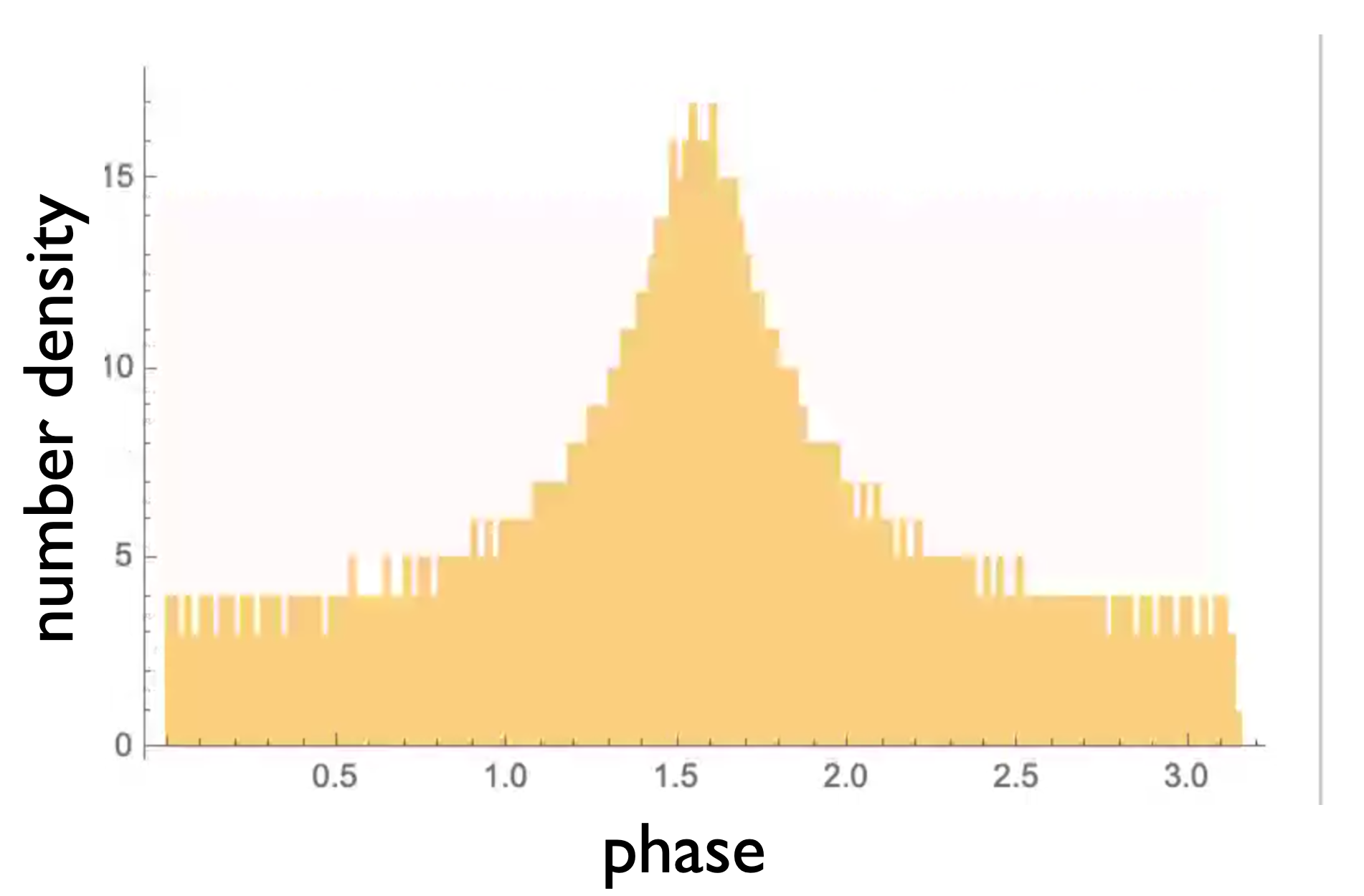}
  \includegraphics[width=.3\linewidth]{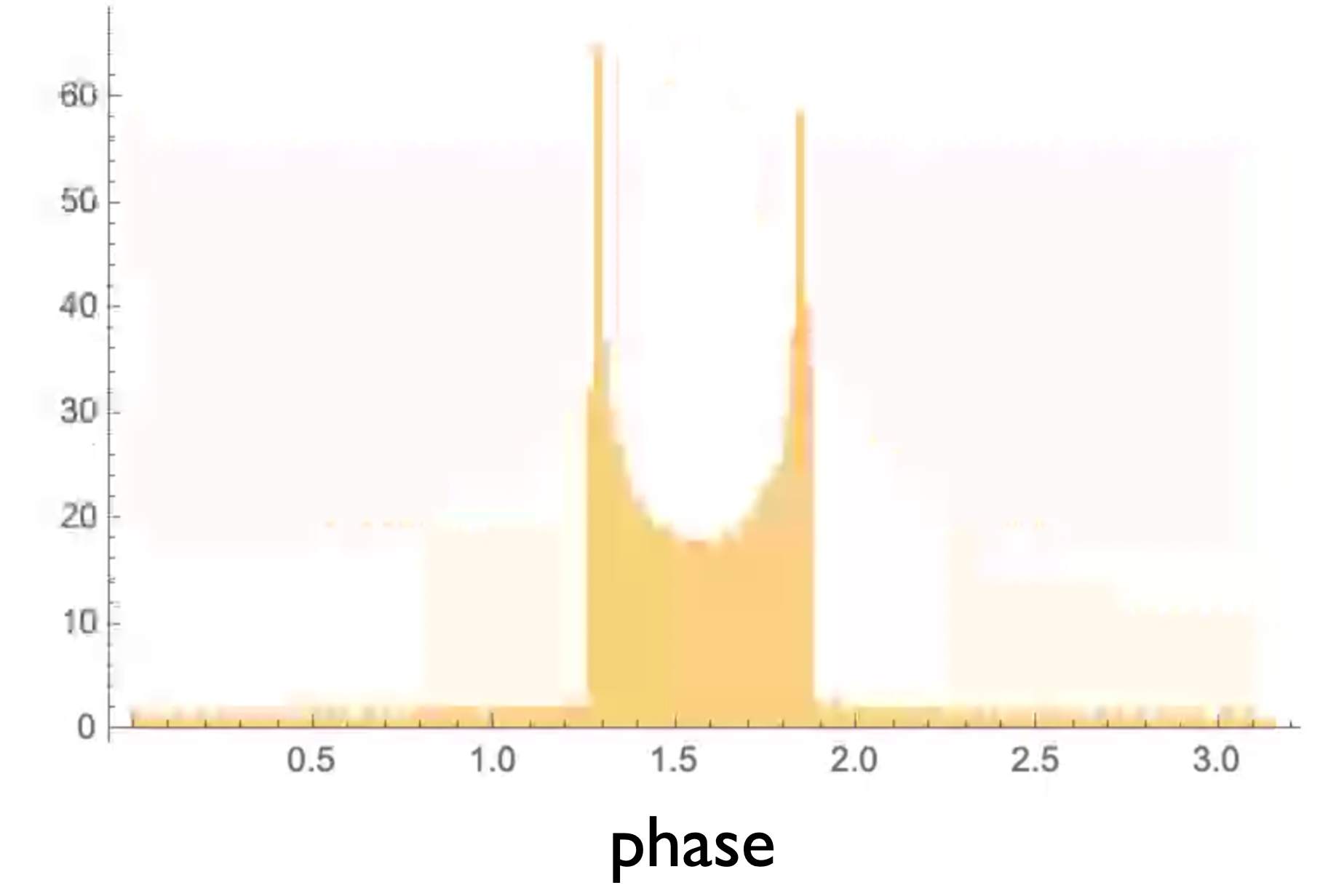}
   \includegraphics[width=.3\linewidth]{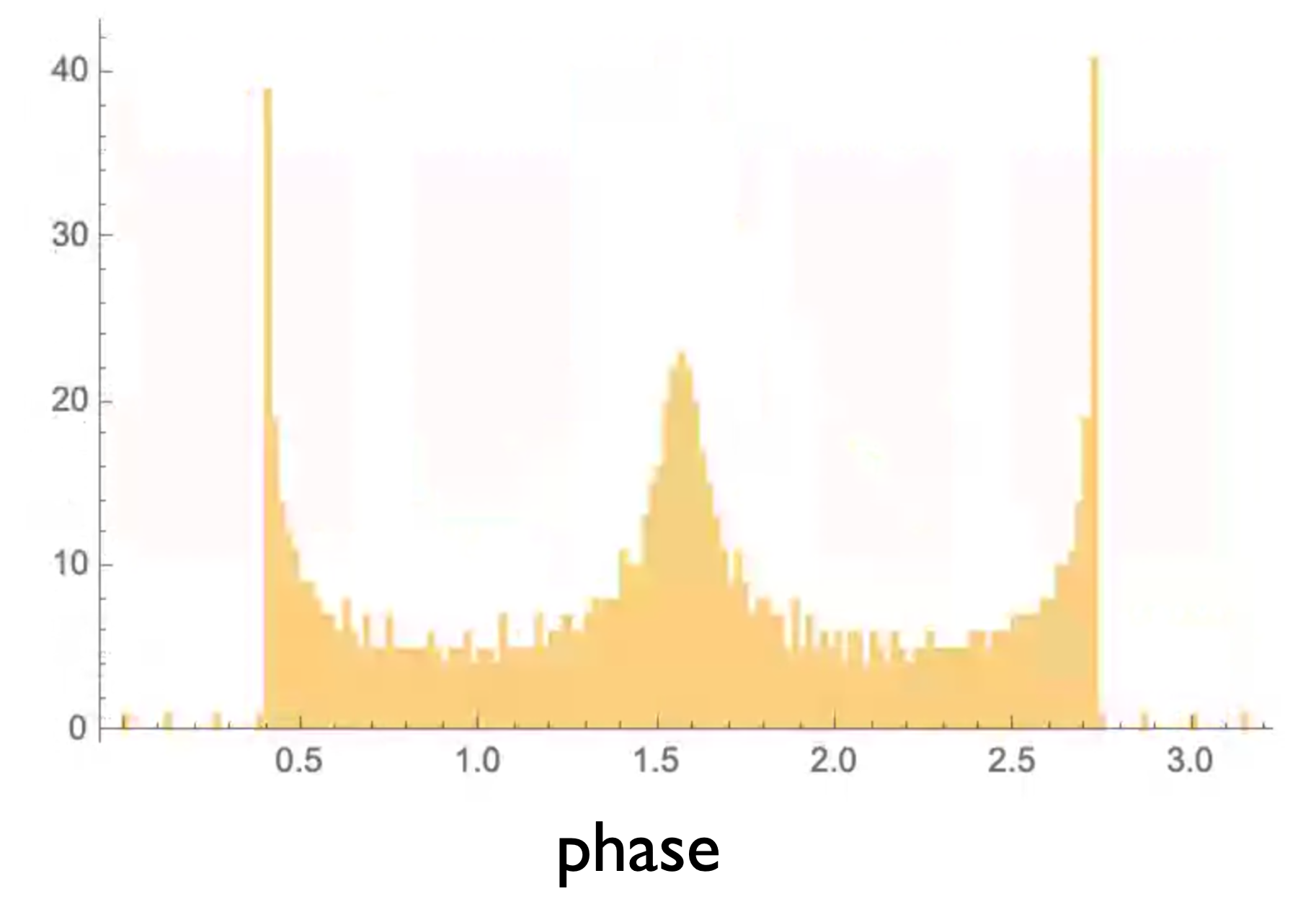}
\caption{Top row. Left panel: Axial particle trajectories for  ${\cal A} \equiv   2  {a_0 a_1 }/({1+a_0^2})  =0.1$ (integration of (\ref{calA})). Initially, particles are homogeneously distributed along $x$ axis (two periods shown). Within each period, particles start osculating, creating double density enhancements. 
Center panel: Phase portrait of particles' oscillations (velocity as function of phase); numerical solution of (\ref{calA}) for  ${\cal A} =  1$. The pattern rotates. With time, the spiral becomes tighter. Eventually, particles will be hot near O-points and cold near X-points. Right panel: Bragg's scattering on induced density bunches.   Two density structures are the result of direct particle integration in the field of two counter-propagating waves with equal intensities of $a_0 = 0.1$. The resulting density structures for pair plasma is logarithmically divergent \citep{2021ApJ...922..166L}. Dashed lines illustrate the coherent addition of the back-scattered wave.
Bottom row: time evolution of density profile within each half wavelength (importantly, the beat pattern is non-stationary/multi-harmonic.}
\label{plus-minus1}
\end{figure}

In Fig. \ref{plus-minus1} we show results of direct  numerical integration of single particles' trajectories  showing formation of two   density bunches per period. Most importantly, at the peak the density is {logarithmically divergent},  see  Ref.  \citep{2021ApJ...922..166L}. Thus, even weak driving with $a_0, \, a_1 \ll 1$ produces in pair plasma divergent density fluctuations. ({We also comment that in pair plasma, in subcritical regime $n \leq n_{cr}$, single-particle trajectories qualitatively match full PIC simulations.})


The dominant mode has  two  bunches per period. 
This is exactly the condition for constructive  Bragg's scattering, top right panel in  Fig. \ref{plus-minus1}. 

\begin{figure}[h!]
  \centering
  \includegraphics[width=.44\linewidth]{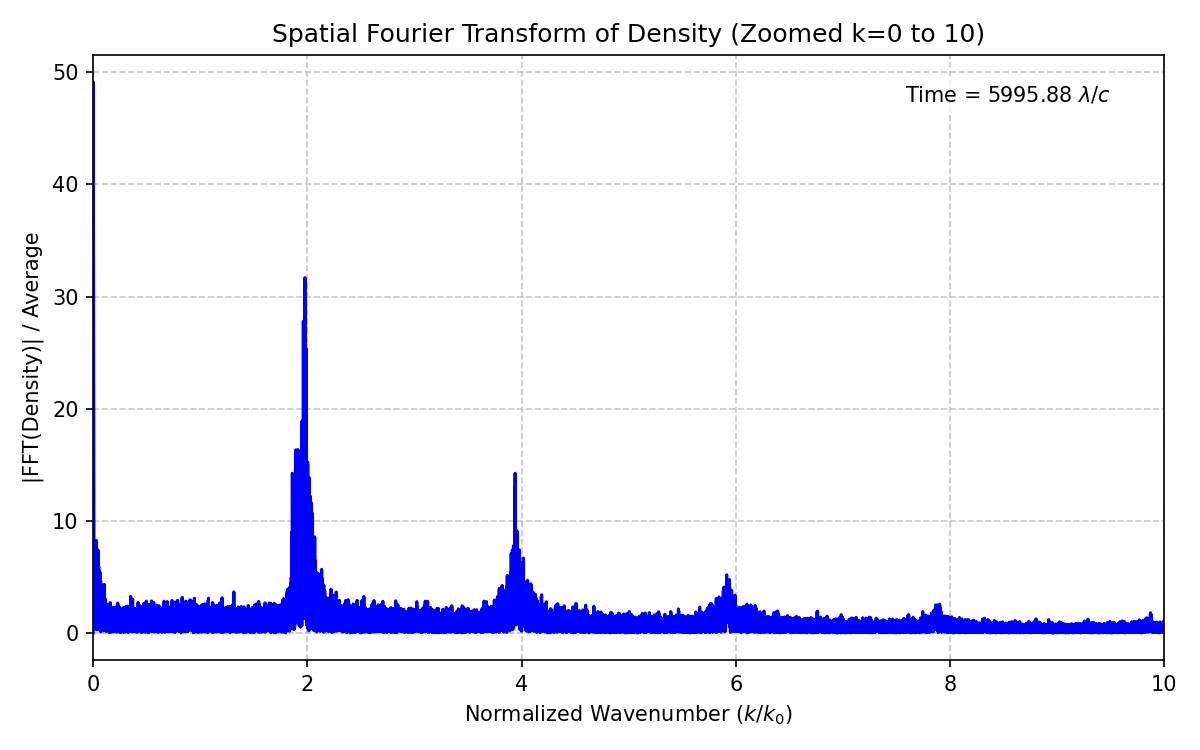}
   \includegraphics[width=.54\linewidth]{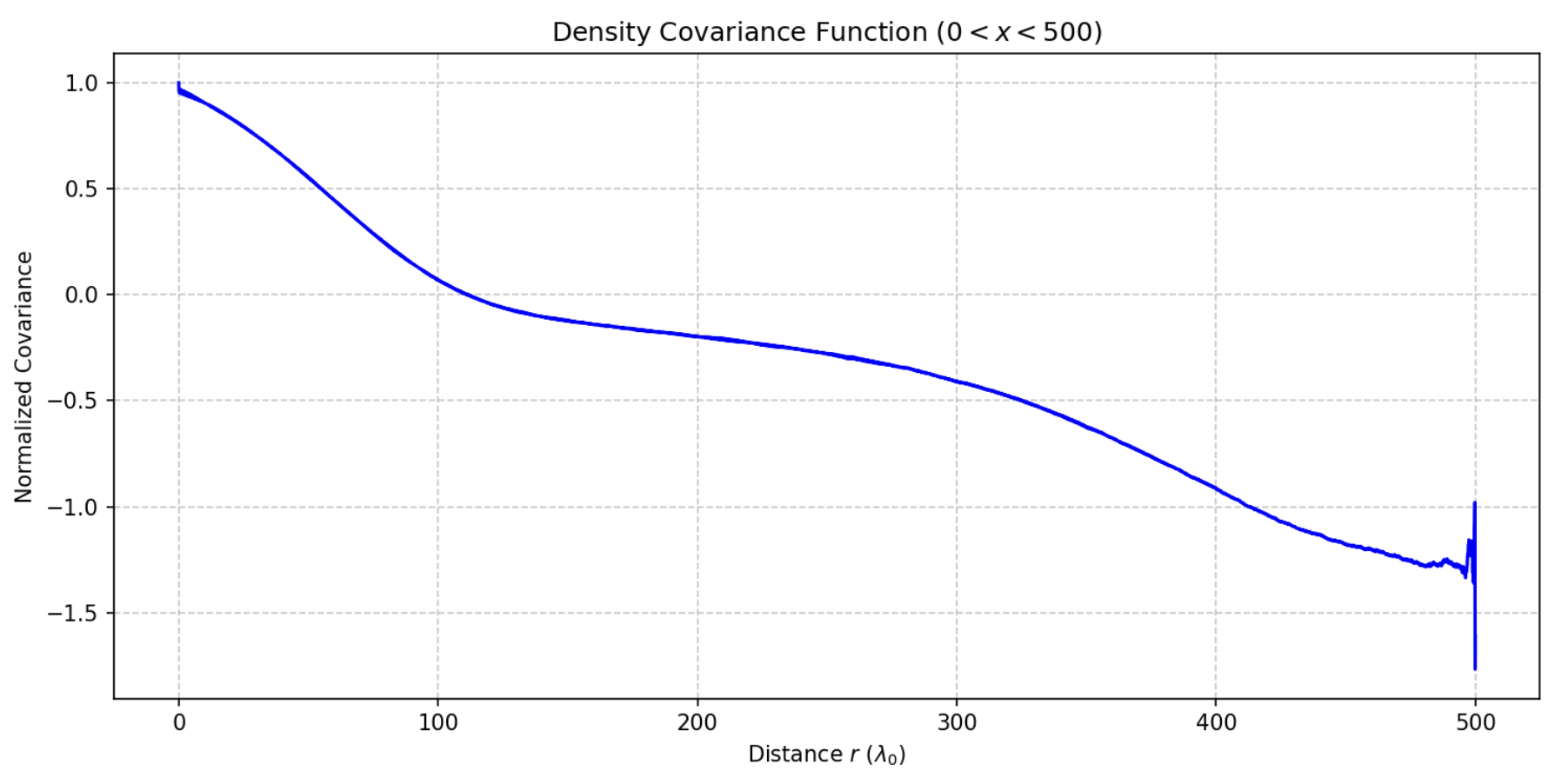}
\caption{  Left: Fourier transform  of  density fluctuations in PIC simulations, zoomed in to  the range $0<k<10k_0$, clearly  showing the peak at Bragg's condition $k= 2 k_0$. Right panel: density covariance (zoomed in to 500 $\lambda$).  The run corresponds to Fig.  \ref{longCP-ep-001}. }
\label{dens0019_fft_zoom}
\end{figure}

{\color{red}

Even in the simple model of two counter-propagating waves, the motion of particles has contributions from several harmonics, creating time-dependent density structures, Fig.  \ref{plus-minus1}, bottom row. 
As a result, a random, space and time-dependent collection of density sheets will be generated. They result in space and time-dependent fluctuations of the dielectric constant. But fluctuations are not completely random - both short and long-scale fluctuations are correlated.  In Fig. \ref{dens0019_fft_zoom}  we plot Fourier transform of the density fluctuations  and density covariance,  clearly showing both  a peak at $2 k_0$  (exactly as expected for  Bragg's scattering)  and long-range correlations. We also note, that the correlations of the \EM\ energy (not shown) occur on scales $\sim 1000 \lambda$, about ten times larger than those of density.}

\subsection{Linear regime of parametric instability}
\label{linear}

\subsubsection{Cold limit}
{Next we consider analytically the linear stage of the parametric instability of a CP wave in pair plasma. The approach generally follows the so-called {\it high gain Compton regime of SASE FELs} \citep
{1950BSTJ...29..608P,1985NIMPA.239....1S,Freund2024}. An important point here is that in pair plasms space-charge effects are absent altogether. 

 There is extensive literature on parametric instabilities in plasmas 
\citep[][]{1974PhFl...17..778D,2016PhRvL.116a5004E,2025arXiv251012869N}. Below we single-out one particularly powerful channel,  specific to pair plasma.

We introduce two unit vectors corresponding to two sences of CP:
\be
{\bf e} _{1,2} = \left\{0, \pm i,1\right\} \frac{1}{\sqrt{2} }
\ee

Let the  initial wave, the driver of frequency $\omega$ and wave vector  $k$, propagating to the right along $x$ direction, be polarized as ${\bf e} _{1}$, while the back-scattered  be polarized along  ${\bf e}  _2$, hence The PLUS mode,
\be
{\bf A}= a_0 {\bf e}  _1 e^{-i ( \omega t -k  x)} +
a_1  {\bf e}  _2 e^{i( \omega ^{(1)} t + k^{(1)} x)}
\ee

In the case of cold plasma, the relations appear especially revealing, The beat between the driver and the back-scattered wave produces nonlinear  axial Lorentz force
\ba &&
d_t v_x = {\bf v} ^{(1)} \times {\bf B} ^{(0)}  +  {\bf v} ^{(0)} \times {\bf B} ^{(1)} =f_x
\nn &&
f_x = \frac{1}{2} i a_0 a_1 \left(k+k_1\right) e^{-i \left(\left(k+k_1\right) x+t
   \left(\omega _1-\omega\right)\right)}
   \label{vx}
\ea
The force on the positive and negative components are the same. As a result the axial motion of both components matches, there is no charge separation.

For $k_1 = k$ (and hence $\om_1 = \om $) the beat force is constant in time
\be
f_x^{(\rm res)}= i a_0 a_1 k e^{-2 i k x}
   \ee
   This indicates that the instability is extremely powerful, as there is no time-averaging.
    In what follows we assume  that these resonant scattering  conditions are satisfied.

 Looking for solution in the form
   \be
   v_x =  {\cal V}(t) e^{-2 i k x} 
   \ee
   we find
   \be
   \dot{ {\cal V}} =- i  a_0  a_1 k
   \label{dotV}
   \ee
   
  The axial  continuity equation for each species reads 
   \be
   \partial_t n + \partial_x (v_x n)=0,
   \ee
 (Here is an important difference from the FEL theory, which deals  with charged beams. In that case, the continuity equation is usually written as
    $
    \partial_t \delta n = \partial_x J_x
    $.
     In our case of pair plasma  $J_x =0$, but  $\delta n \neq 0$.)
  
 Writing  perturbations in the form
   \be
   n = n_0 (1+ f(t)  e^{-2 i k x}  ),
   \ee
   and expanding in small $f$, we find
   \be
   \dot{f} =  2 i k  {\cal V}
   \label{dof}
   \ee
   
   Finally, in  the non-linear part of the wave equation for the amplitude of the back-scattered wave   $a_1 \equiv a_1 (t)$, we  drop the non-resonant current term $\propto a_1$, and the fast,  spatially oscillating  terms. We find
   \be
   \om  \partial_t a_1 = i a_0  f(t) \om_p^2
   \label{a1}
   \ee

   Equations (\ref{dotV}),  (\ref{dof})) and (\ref{a1}) form a complete set that describes the nonlinear evolution of the back-scattered wave. The  linear system can be reduced to
   \be
   i \om  \partial^3_t a_1 + 2 a_0^2 k^2 \om_p^2 a_1 =0
   \ee

   Looking for a solution $\propto e^{\Gamma t}$,
   \be
   i \Gamma ^3 \omega+ 2 a_0^2 k^2 \omega _p^2 =0
   \ee
   Since $ \om  \approx k$ (for $\om  \gg \om_p$),
   we find the growth rate   (subscript $c$ indicates the cold plasma limit) 
   \be
    \Gamma_c  \equiv  \frac{\sqrt{3}}{2^{2/3}} \left( a_0^2 \omega \omega _p^2 \right) ^{1/3} = \frac{\sqrt{3}}{2^{2/3}}    \rho_L \om
    \label{Gammac}
\ee
where we defined 
\be 
   \rho_L=     \left(    a_0 \frac{  \om_{p} }{\om }\right)^{2/3}  
   \label{rhoL}
     \ee

      Qualitatively, the result (\ref{Gammac}) is exactly what is expected. The quantity  $c/\Gamma$ is  what is called gain length in the Self-Amplified
Spontaneous Emission (SASE) regime of free electron lasers (FEL).
Parameter $\rho_L $introduced in (\ref{rhoL}) is approximately 
 the Pierce parameter of FEL  expressed in terms of properties in the beam frame \cite{1950BSTJ...29..608P,Freund2024}. Growth rate (\ref{Gammac}) 
  also matches the rate for strongly coupled Brillouin scattering,  Eq. (12) of \cite{2016PhPl...23h3122E}, with a substitution of electrons plasma frequency instead of the ion one. \citep[In a related work, in Ref. ][ enhanced seeded, not self-generated, Brillouin Backscattering was considered]{2016PhRvL.116a5004E}.

   The corresponding  instability length (the analogue of the FEL  gain length) is
   \be
   L=  \frac{c}{\Gamma }=   \frac{1}{\rho_L} \frac{c}{\om}
   \propto a_0^{-2/3} \left(\frac{n}{n_{cr}}\right) ^{-2/3} 
   \label{L}
   \ee
where the latter scalings are to be confirmed with numerical simulations, \S \ref{PICC}. Scale  (\ref{L}) is the estimate of the length of the ``nose'' that passes through. 


Finally, the growth rate (\ref{Gammac}) should be smaller that a typical decoherence rate due to jitter velocity of particles in the wave $\sim  a_0 \om$. This requires $\rho_L \leq a_0$,
\be
a_0 \geq \frac{\om_p^2}{\om^2} \equiv \frac{n}{n_{cr}}
\label{akh}
\ee

\subsubsection{Thermal effects}

The main modifications in warm plasma (but still in  a fluid regime) is that Eq.  (\ref{vx})  should now be written not for a single particle, but for each fluid. Pressure forces are expected to counteract density compression. Assuming for simplicity isothermal plasma with sound speed $c_s$ (normalize to the speed of light). The longitudinal force-balance equation now becomes
\ba &&
\partial_t v_x= f_x^{(\rm res)} -  c_s^2 \partial_x \ln n
\nn &&
 \dot{ {\cal V}} =
 -i k \left(a_0 a_1-2 c_s^2 f(t)\right) 
 \ea
 
 The growth rate obeys
 \be
2 a_0^2 \omega \omega _p^2+4 i \Gamma  \omega^2 c_s^2+i \Gamma ^3=0
 \ee

There is a typical sound velocity (and correspoding temperature) 
when  real part of $\Gamma$ (responsible for the growth) formally becomes zero at
\be
c_s^\ast = \frac{\sqrt{3}}{2}   \left(    a_0 \frac{  \om_{p} }{\om }\right)^{2/3}  \approx \rho _{L}\
\label{cs0}
\ee
(times the  speed of light).


   For smaller $c_s \leq c_s^\ast$, the instability growth rate in warm plasma $ \Gamma_w$ is
   \be
    \Gamma_w =
   \Gamma_c \left( 1 - \left( \frac{c_s}{ c_s^\ast } \right) ^2 \right) \approx \Gamma_c \left( 1 - \frac{\Theta}{\rho_L^2} \right) 
   \label{Gsm}
    \ee

For $c_s \geq  c_s^\ast $ the parametric instability enter the kinetic regime. In this case  the growth rate is \cite{2022ApJ...930..106G,2025arXiv251012869N}
\ba && 
\Gamma_h \approx a_0^2 \frac{\om_p^2}{\om }  \frac{1}{\Theta} \approx \frac{\rho_L^3}{\Theta} \om
\nn &&
\Theta = \frac{k_B T}{m_e c^2}
\label{Gammah}
\ea

For given $a_0$ and $\om $, the kinetic and cold hydrodynamic growth rates are equal at 
\be
\Theta_{h-k}=  \left(    a_0 \frac{  \om_{p} }{\om }\right)^{4/3} = \rho_L^2
\label{Thetakh}
\ee
(see also Ref.  \cite{2022ApJ...930..106G}, Eq. (30), and  Ref. \cite{2025arXiv251012869N}, Eq. (50)).
The condition (\ref{akh}) then ensures that  $\Theta_{h-k} \leq a_0^2$. 
}

\subsection{PIC simulations of the linear stage, temperature-induced transparency,  and seeded back-scattering}
\label{PICC}

\subsubsection{Scaling of the ``nose'' part with $a_0$ and density}

{
Our PIC simulations confirm these theoretical estimates. First, there is only one reproducing feature of the simulations: the length of the "nose". We associate it with the instability length (\ref{L}). 
 In Fig. \ref{brillion-localiz} we compare the length of the ``nose'' for various $a_0$ and $n/n_{cr}$. As expected, for  higher $a_0$ and denser plasmas, the instability length decreases.
  \begin{figure}[h!]
  \centering
 \includegraphics[width=.49\linewidth]{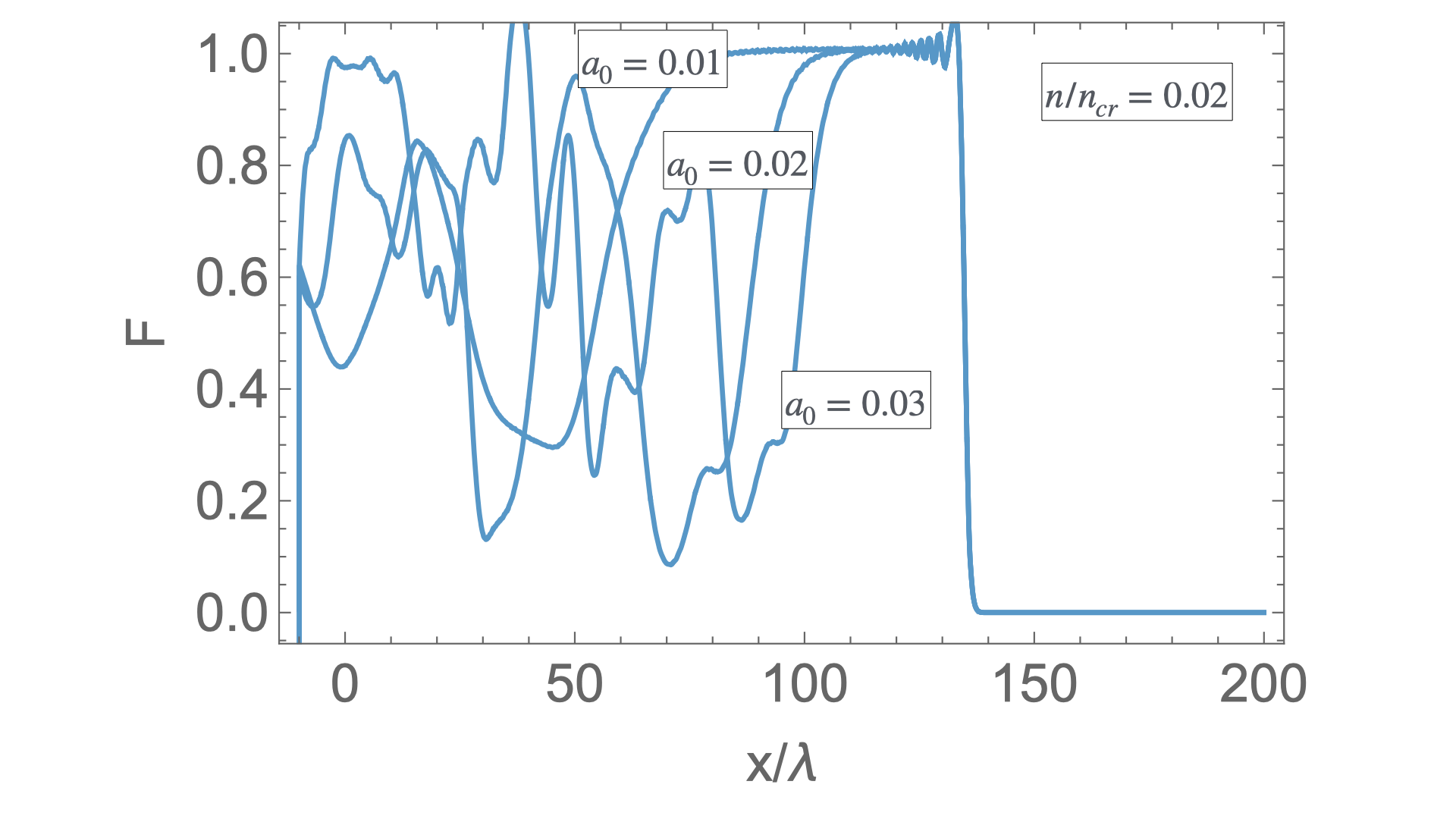}
 \includegraphics[width=.49\linewidth]{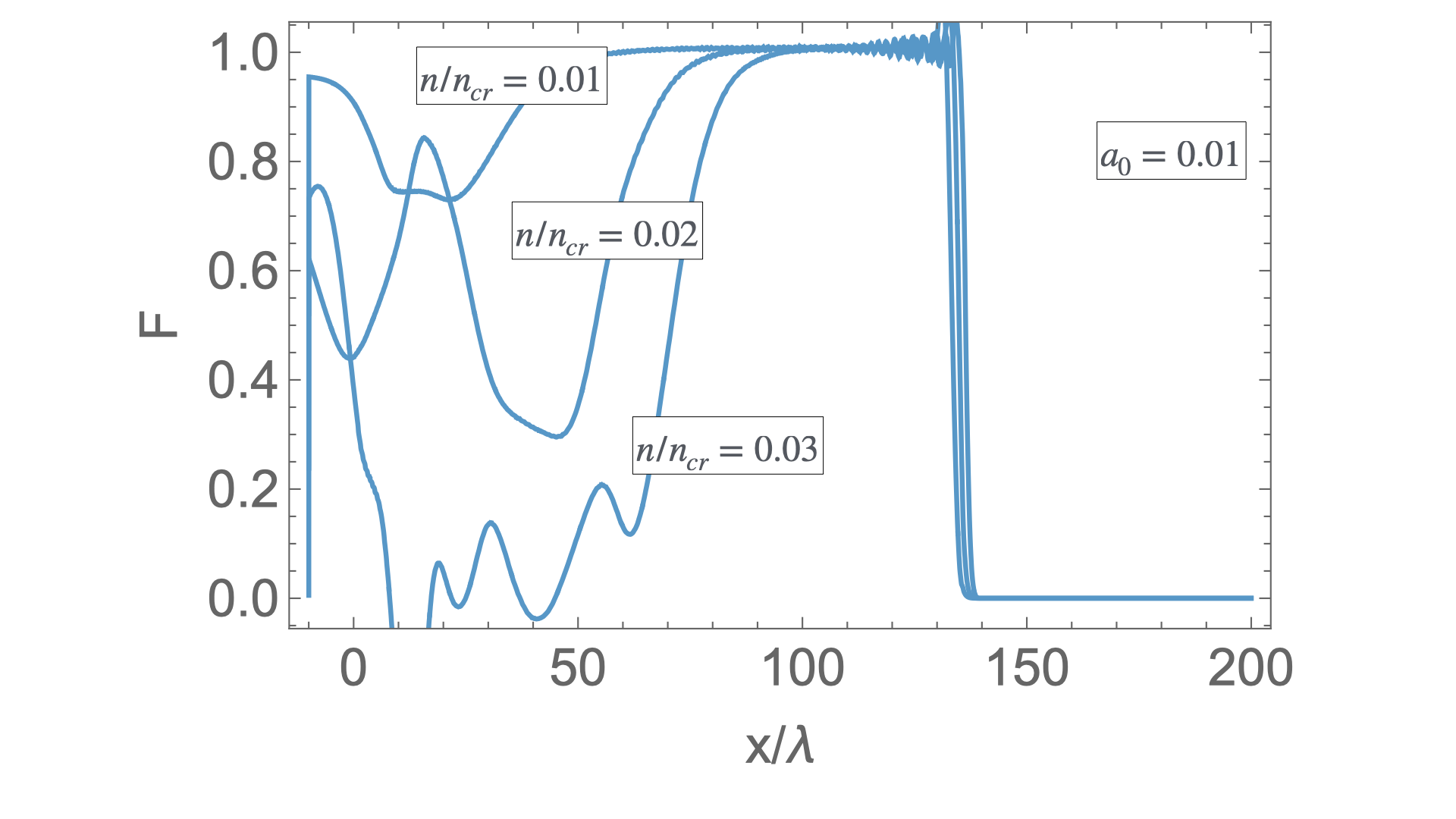}
\caption{ Comparing the expectations from linear theory, Eq. (\ref{L}) with numerical runs for fixed density (left panel) and fixed intensity (right panel). Results are qualitatively consistent with expectations (shorter instability length for higher $a_0$ and denser plasma).}
\label{brillion-localiz}
\end{figure}
}

\subsubsection{Temperature-induced  transparency}
\label{Tstabilization}

\begin{figure}[h!]
  \centering
 \includegraphics[width=.99\linewidth]{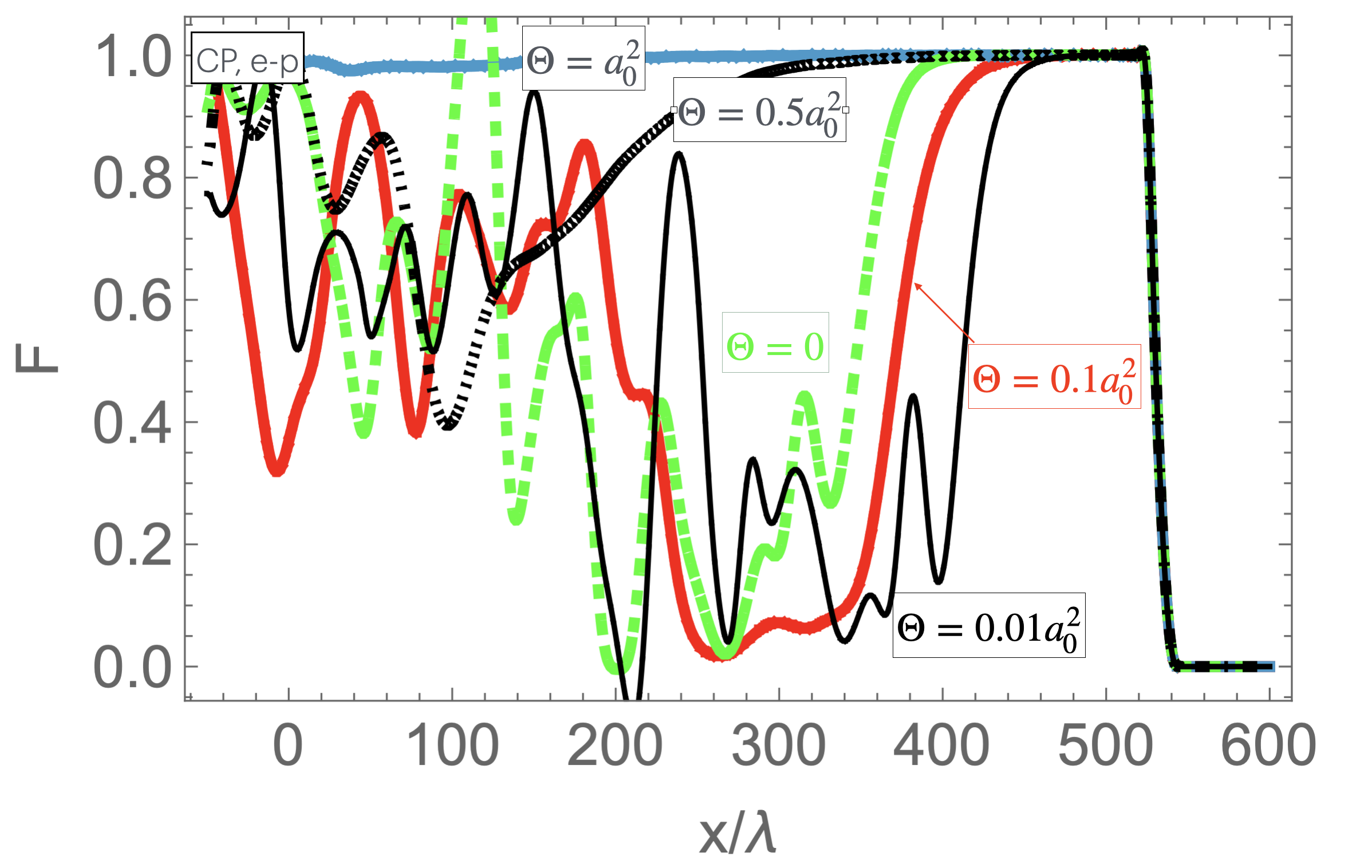}
\caption{Temperature-induced  transparency (plotted is Poynting flux for CP, $e^\pm$ plasma, $a_0 =10^{-2}$). {Low temperatures $\Theta \leq 0.1 a_0^2$ (green dashed, red, solid black curves) do not affect wave propagation much}. For  $\Theta \geq 0.5 a_0^2$ (black dot-dashed and solid blue curves) transparency is restored. }
\label{T0}
\end{figure}

Beyond some limit, the initial temperature has the most drastic effect on the transparency of pair plasma,  Fig. \ref{T0}. Qualitatively,  the jitter motion of a particle in a wave corresponds to temperature 
\be
\Theta_{a_0} \equiv T/(m_e c^2) \sim  a_0^2 
\label{Thetaa0}
\ee 
When the plasma temperature approaches $\Theta_{a_0} $,   the properties of wave propagation are expected to be strongly modified compared to the cold case. Our simulation conforms with this expectation, Fig. \ref{T0}:  for $\Theta \leq \Theta_{a_0} $, the behavior is very similar (with some variations) to the cold case. At higher temperatures,  $\Theta \sim \Theta_{a_0} $, transparency is restored; see also \cite{2025arXiv250906230T} and Appendix \ref{Epoch}.

The reason for temperature-induced  transparency is that a
 relatively 
 small temperature  spread  destroys the initial seeds of parametric instability of density bunches (the beat of the driver and back-scattered wave) that eventually lead to the creation of large density fluctuations.

Temperature-induced  transparency adds another level of complexity to the problem: forward parts of the pulse warm-up the initially cold plasma. This in turn suppresses the density seeds of the parametrically-induced density perturbations, making plasma more transparent. The results for longer runs, Fig. \ref{longCP-ep-001}, indicate that plasma heating by the pulse  is not sufficient  in the case $a_0=10^{-2}$ to make plasma transparent.

\subsubsection{Seeded back-scattered wave}

\begin{figure}[h!]
  \centering
 \includegraphics[width=.99\linewidth]{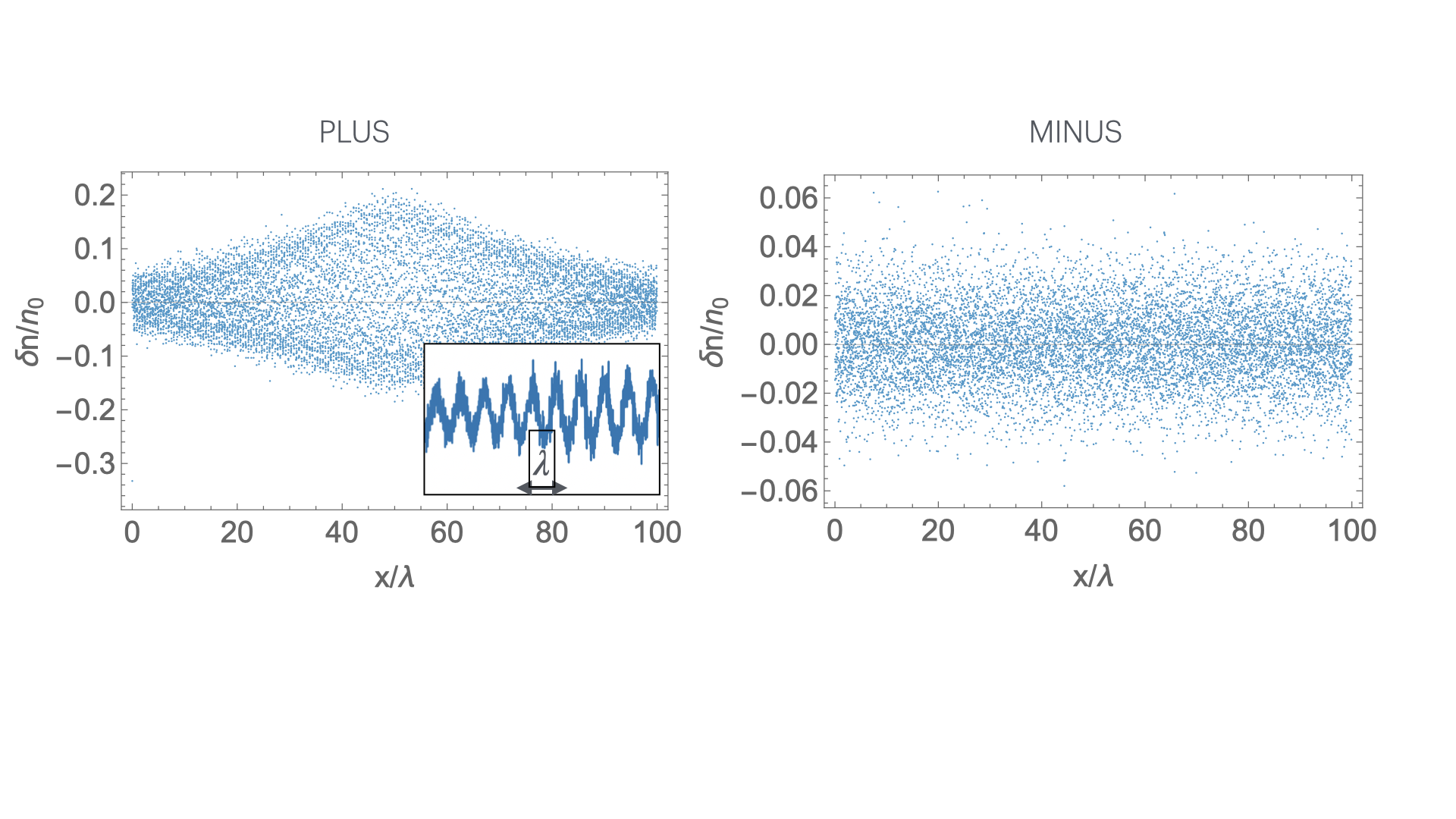}
 \includegraphics[width=.99\linewidth]{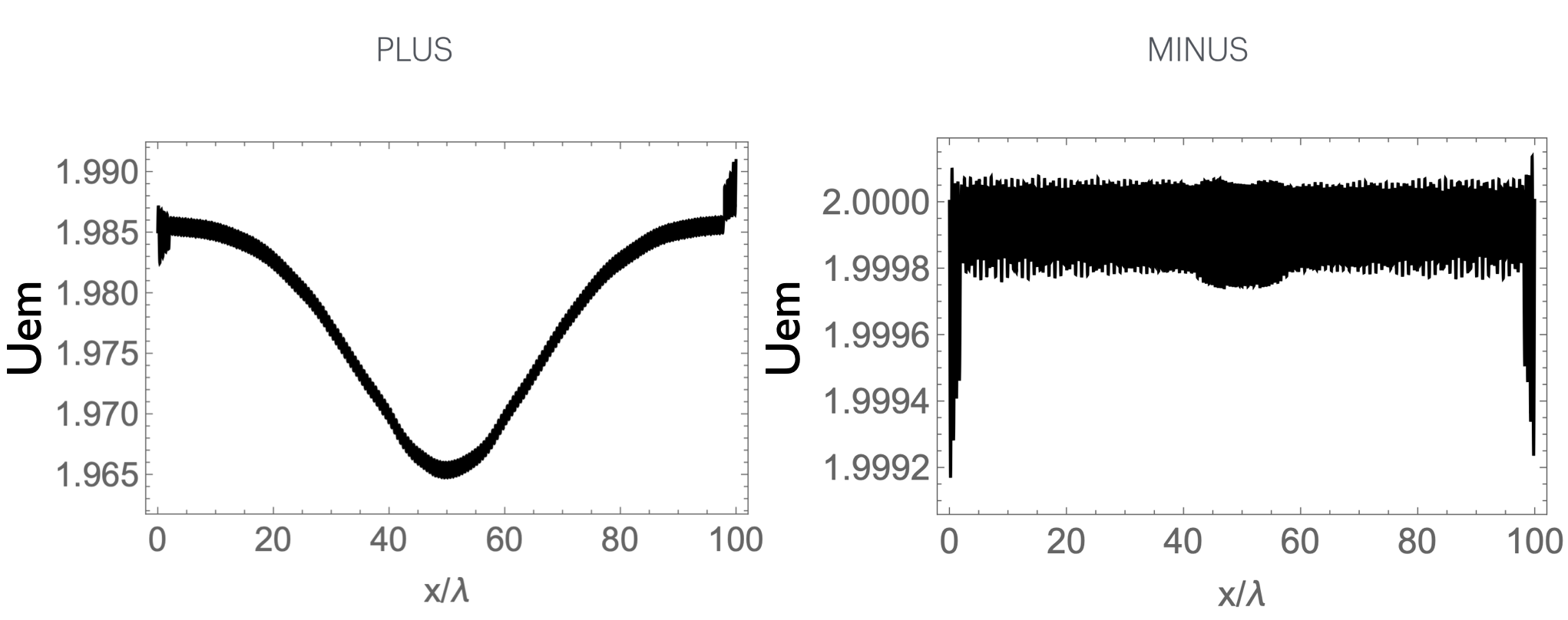}
\caption{Seeding of density fluctuations. Top row: Density fluctuations for two CP waves launched from the opposite sides of the simulation box for two polarizations; $a_0 =10^{-3}$. Notice different scale in the PLUS and MINUS plots. As expected, density fluctuations in the PLUS configuration are much larger (level of density fluctuations in the MINUS configuration is consistent with initial numerical noise. Bottom row: energy density of EM fields. Notice different scales. { The insert shows a zoomed-in view of density in the PLUS configuration, showing large modulation at half wavelength $\lambda/2$}.} 
\label{PIC-plus-minus}
\end{figure}

In a complementary run, we ``seed'' the modulational process by launching a weak back-scattered wave. 
In  Fig. \ref{PIC-plus-minus} we plot density fluctuations for two CP waves launched from the opposite sides of the simulation box for two relative polarizations. This configuration effectively seeds the density fluctuations, after the waves penetrate each other. We observe that in the PLUS configuration density fluctuations are an order of magnitude higher, as expected. 
This result also serves as a test of the code, confirming that the code produces correct results for $a_0$ as small $ 10^{-3}$. (The code treats the outgoing wave effectively as an open boundary.)

We also observe effects of weak localization: little dip in the center, left panel bottom row in Fig. \ref{PIC-plus-minus}. In the high density spikes PLUS configuration, energy density in the center is smaller by $\sim$ few $10^{-2}$ (values normalized). We interpret the dip as weak reflection in both directions,  due to weak localization by density structures.  In the MINUS configuration the energy density is constant to $\sim 10^{-4}$.   

{
Thus, in the beat of two counter-propagating waves, large density perturbations are created. Most importantly, {\it in pair plasma density bunches are charge neutral}. Thus, there is no 
electrostatic repulsion \citep[this is different from the SASE regime of FELs; see also a similar  numerical experiment on Plasma Photonic Crystals  in electron-ion plasma][] {2003ApPhB..77..673S,2016PhRvL.116v5002L,10.1063/5.0096386}. }

\subsection{Summary of important parameters}

The problem we consider is highly nonlinear, with a number of parameters involved. 
Let us  summarize different regimes. There are three dimensionless parameters in the system: laser intensity $a_0$, ratio of frequencies $\om_p/\om  \approx \sqrt{n/n_{cr}}$, and plasma temperature $\Theta = k_B T/(m_e c^2)$ (polarization is less important). These parameters appear in various  combinations:
\begin{itemize}
    \item Pierce-like parameter $\rho_L$ (\ref{rhoL}). It determines the linear stage of the instability in the cold regime
    \item  Ratio 
    \be
    \eta_L= (a_0/\rho_L)^3 = a_0 \om^2/\om_p^2 \equiv \left(\frac{\Theta_{a_0}}{\Theta_{h-k}}  \right)^{3/2},
    \label{eta}
    \ee Eq. (\ref{akh}). For $\eta \geq 1$ the range for the kinetic regime of the instability disappears.
    \item Combination $a_0 \om/\om_p$, Eq. (\ref{a0c}), approximate the  transition from ponderomotive to snowplow regime 
    \item  $\Theta_{a_0} = a_0^2$, Eq. (\ref{Thetaa0})  - temperature of  full transparency
    \item  $\Theta_{h-k} = \rho_L^2$, Eq. (\ref{Thetakh}) -  transition of linear stage of the parametric instability with growth rate $\Gamma_c$, Eq. (\ref{Gammac}),  into kinetic regime with growth rate $\Gamma_h$, Eq. (\ref{Gammah}).
\end{itemize}


\section{Localization of an EM wave already inside pair plasma: the wave  equation approach}
\label{inside}

Above in \S \ref{Reflection} we considered one aspect of Anderson localization of EM waves: reflection by an under-dense pair plasma. Next, we address a somewhat different but related question: if a wave is already present inside plasma, generated in the regime where no localization occurs (\eg\ in regions of high guide \Bf), it then propagates, while still inside plasma, into regions where localization does happen.

The following processes will occur.  Initially, the wave occupied all space;  it was carrying energy and momentum. After the scatterers have grown, the wave cannot propagate, it will become localized, separating into bright regions (occupied by standing waves)  with low plasma density and dark regions with high plasma density. The initial  momentum of the wave will be transferred to the material. Effectively, the EM wave stops (in plasma frame).
This behavior resembles slowing down of waves in  disordered photonic time crystals \cite{2021PhRvL.126p3902S}.

We model the disorder as a collection of delta-thick scatterers. For consistency,  in Appendix \ref{deltaspike} we consider stationary case: reflection of an external wave, by a stationary collection of delta-scatters.

Below, in \S \ref{Wave-disorder} we introduce the method, assuming stationary scatterers. Later  in \S \ref{swithc}
we consider slow  (adiabatic, on time scales much longer than wave period) switching-on of the scattering  ``walls''.



 \subsection{Wave equation with random delta-disorder}
 \label{Wave-disorder}

Here we first  recast the problem of Anderson localization of waves  to the problem of waves inside plasma with already present delta-fluctuations of the permittivity. 

{Let us specialize to a wave equation describing an electromagnetic wave propagating along the $x$-direction with the electric  permittivity varying in the same $x$-direction while being independent of other coordinates. In this geometry, the electric field can be chosen to point in the $y$-direction, while the magnetic field points in the $z$-direction. Both can be parametrized by a gauge potential $A=A_y$ pointing in the same direction as the electric field,
that is 
 in the $y$ direction. 

The disordered wave equation at a given frequency $\omega$ then takes the form
\be 
\d^2_x A - \epsilon(x) \, \d^2_t A=0,
\ee
where the electric permittivity $\epsilon(x)>0$ is a random function of position.  }
In the absence of disorder $\epsilon=1$.
Just as elsewhere we set the speed of light $c=1$.

We write
\be \epsilon(x) = 1 - V(x),
\ee
where $V(x)$ represents disorder.
This now gives in the frequency domain
\be 
\label{eq:dw} 
 -\d_x^2 A + \omega^2 V(x)  A
 =\omega^2 A.
\ee
In this form, the disordered wave equation 
resembles the Scr\"odinger equation with random potential quite closely. 
One obvious difference is that the disorder $V(x)$ gets multiplied
by the square of the frequency $\omega$. As a result, low frequency suppresses
disorder, as was already discussed earlier. 

It has been known for quite some time
that in one-dimensional space all
solutions of Eq.~(\ref{eq:dw}) are localized
\cite{GoldsheidMolchanovPastur1977,LeeRamakrishnan1985,EversMirlin2008},
which means they have the form
\be A(x) \sim f(x) \, e^{- \frac{\left| x-x_0 \right|}{{\cal L}}},
\ee where ${\cal L}$ is called the localization
length and $x_0$ is the center of localization,
and $f(x)$ is a randomly oscillating function
of position $x$ whose magnitude is 
of the order of 1. A particular
well established result in the theory
of disordered wave equations states that the localization length of the 
disordered one dimensional wave equation such as (\ref{eq:dw}) diverges with the frequency
of the wave as ${\cal L} \propto 1/\omega^2$ \cite{JohnSompolinskyStephen1983,GurarieChalker2003}.

Next, we represent our disorder $V(x)$ as a collection of delta-scatterers:
\be V(x) = \ell \sum_{i=1}^N \delta(x-x_i).
\ee
Here $\ell$ is a parameter with the dimension of length. $x_i$ will be taken as placed in random positions. 

The average potential is
\be \label{eq:avv} \VEV{V(x)} = \ell \sum_{i=1}^N \int_{-L/2}^{L/2} \frac{dx_i}{L} \delta(x-x_i) = \ell \frac{N}{L} = \frac{\ell}{a},
\ee
(where $a=L/N$ is the average distance between $x_i$), while the variance is
\be \label{eq:corv} \VEV{V(x) V(y)} = \ell^2 \sum_{i=1}^N \int_{-L/2}^{L/2} \frac{dx_i}{L} \delta(x-x_i) \delta(y-x_i) + \ell^2 \sum_{i \not = j} \int_{-L/2}^{L/2}\frac{dx_i dx_j}{L^2} \delta(x-x_i) \delta(x-x_j) =
\ee
\be =  \ell^2 \frac{N}{L} \delta(x-y) + \ell^2 \frac{N(N-1)}{L^2} \approx \VEV{V(x)} \VEV{V(y)} + \frac{\ell^2}{a} \delta(x-y). 
\ee

It is very well known that the localization
length of the solutions of disordered Schr\"odigner equations in one dimension
is of the order of the mean free path
\cite{Thouless1973}. At weak
disorder strength, which in our case
of the wave equation is synonymous to low frequency $\omega$, it has been further
suggested that the localization length is
twice the mean free path. Mean
free path is the average distance a wave 
propagates before it scatters off the disorder.
Mean free path is relatively easy to calculate,
especially at weak disorder strength where it
can be calculated perturbatively.

Let us determine mean free path of our wave
equation at low frequency $\omega$. 
The structure of the wave equation \rf{eq:dw} allows
us to us perturbation theory since at small $\omega$ disorder is effectively weak. 
We will rely on the standard method of Green's functions. 
First we define the retarded Green's function
in the position domain
\be G = \left[ \omega^2 - \omega^2 V(x) + \d_x^2 \right]^{-1}. 
\ee
Averaged over disorder, the Green's function
in the momentum domain is given by 
\be G = \frac 1 {\omega^2 -   k^2 - \Sigma(\omega)},
\ee
where $\Sigma(\omega)$ is the self-energy.
If we know self-energy, the strategy for calculating the mean free path consists of
finding the pole $\omega_p$ of the Green's
function satisfying
\be \omega_p^2 -   k^2 - \Sigma(\omega_p)=0.
\ee
An imaginary part of the pole gives the scattering rate, which can then be converted 
into the mean free path by dividing the speed 
of light  (set to be $1$ in this calculation) by this rate. 

We now need to determine 
the self-energy $\Sigma$. 
At weak disorder strength $\ell$ or at low frequency $\omega$, the self energy can be calculated perturbatively \cite{LeeRamakrishnan1985}. Up to the second order of perturbation theory, the self-energy is schematically given by
\be \Sigma = V + VGV + \dots.
\ee
Averaging
over random $V$ by using the relations \rf{eq:avv} and \rf{eq:corv} gives 
\be \Sigma = \omega^2 \frac{\ell}{a} + \frac{\ell^2 \omega^4}{a} \int_{-\infty}^\infty \frac{dq}{2\pi} \frac{1}{(\omega+i0)^2 -   q^2}.
\ee
We are only interested in the imaginary part of $\Sigma$, which is given by
\be {\rm Im} \, \Sigma =-i \pi \frac{\ell^2 \omega^4 }{a}  \int_{-\infty}^\infty \frac{dq}{2\pi} \delta( \omega^2 -   q^2)  =-i \frac{\ell^2 \omega^3}{2   a} 
\ee
We now calculate the scattering rate $\tau^{-1}$ by setting
\be \left( \omega - i \tau^{-1} \right)^2-  k^2 +i \frac{\ell^2 \omega^3}{2  a} =0.
\ee
Solving this in the approximation where $\tau^{-1} \ll \omega$ produces $\omega=k$ and
\be \tau^{-1} = \frac{\ell^2 \omega^2}{4 a}.
\ee
Mean free path is, at small $\ell$,  the speed of light (again, taken to be $1$) multiplied by $\tau$, which gives
\be \xi = \frac{4   a}{\ell^2 \omega^2}.
\ee
The localization length in one dimension, 
at weak disorder strength which in the
current context is enforced by a small frequency
$\omega$,
is equal to twice the mean free path.
Therefore, we find that at low frequency $\omega$ the localization length behaves as
\be
{\cal L} = 2 \xi = \frac{8   a}{\ell^2 \omega^2} = \frac{2 a \lambda^2}{\ell^2 \pi^2}.
\label{calL}
\ee
Here we replaced the frequency $\omega$ with the
wavelength $\lambda$ of the wave according to
$\lambda=2 \pi/\omega$ (again, in the units where the speed of light is $1$). 
Note that its frequency dependence is ${\cal L} \sim 1/\omega^{2}$ in agreement with the 
established results in the theory of disordered wave equations discussed earlier.

For the expected distance between scatters $a \sim \lambda$, and dimensionless strength $\Gamma = \ell/(2 \lambda)$, the expected localization length is $\sim $ few $\lambda$. This is generally consistent with the results of PIC simulations, \S \ref{Clemmow1}.

\subsection{Wave equation with disorder which adiabatically turns on }
\label{swithc}

Here we    analyze the disordered wave equation,
where disorder slowly (adiabatically) turns on.
{Adiabatic switch-on of perturbations is justified by the fact that the parameter $\rho_L \ll 1 $, Eq. (\ref{rhoL}) is small.}
Our goal is to understand how a plane wave, as disorder increases in time,
breaks into localized lumps. Qualitatively, we expect that a plane wave, being an eigenmode of the wave equation without disorder, will adiabatically evolve into an eigenmode of the disordered wave equation. We show below that as this is happening, the amplitude of the wave behaves as a harmonic oscillator whose frequency adiabatically evolves in time. In particular, the number of ``photons", or the ratio of the energy of the wave to its frequency, remains approximately conserved. 

We begin with the action describing a scalar field $A$, representing {the $y$-component of the
gauge potential as explained above}, 
where the electric permittivity changes in time,
\be S= \oh \int dx dt \left[ \epsilon(x,\eta(t)) \left( \partial_t A \right)^2 -   \left( \partial_x A \right)^2 \right].
\ee
Here $\epsilon(x, \eta(t))$ is a random function of $x$ which also depends on time via a parameter $\eta$ which is slowly turned on,
ranging from $0$ at large negative times to $1$ at large positive times. 
For example, it could be that 
\be \epsilon(x,\eta) = 1 - \eta V(x),
\ee where $V(x)$ is a random function of $x$. 
The equation of motion that corresponds to this action is
\be \label{eq:wave} - \d_t \left( \epsilon \d_t A \right) +   \d_x^2 A = 0,
\ee
To understand the solutions to this equation, we consider first a situation where $\eta$ is time-independent, hence so is $\epsilon$. Let's introduce the functions $\psi$ satisfying
\be \label{eq:eigen1} \omega^2 \epsilon \psi = -   \d_x^2 \psi.
\ee
There are many solutions of this eigenstate-style equation, 
producing the spectrum of $\omega$ and the associated eigenstates $\psi$, each dependent on its $\omega$.
Each solution smoothly depends on $\eta$, in the sense of both $\psi$ and $\omega$ being $\eta$-dependent. As we
already established, the solutions of these equations are all localized, unless there is no disorder or $\eta=0$.

Let us normalize $\psi$ by requiring that
\be \label{eq:norm} \int dx \, \epsilon \psi^2=1.
\ee

Now suppose $\eta$ varies in time. We look for a solution to the equation of motion \rf{eq:wave} with the following adiabatic ansatz
\be \label{eq:eigen} A = u(t) \, \psi. 
\ee
Precise criteria exist in the literature regarding the validity of this ansatz. In particular, the rate of change of $\eta$ must be lower than a typical rate controlled by the gaps in the spectrum of $\omega$, in order for this ansatz to be valid. For the analysis here we will just assume that the rate of the change of $\eta$ is sufficiently small, without trying to quantify this further.

Substituting this into the equations of motion and keeping terms no more than  linear in $\dot \eta$ (assuming adiabaticity) we find
\be -\d_t \left[ \epsilon \left( \dot u \psi + u \pbyp{\psi}{\eta}{ \dot \eta}  \right) \right]+   u \d_x^2 \psi =0.
\ee
This can be brought to the following form
\be - \pbyp{\epsilon}{\eta} \dot \eta \, \dot u \, \psi - \epsilon \ddot u \psi - 2 \epsilon \dot u \pbyp{\psi}{\eta} \dot \eta  - u \, \omega^2 \epsilon \psi=0.
\ee
We now multiply this equation by $\psi$ and integrate over $x$, taking advantage of \rf{eq:norm}. We find
\be - \ddot u - \dot \eta \dot u  \int dx \pp{\eta} \left( \epsilon \psi^2 \right) -  \omega^2 u =0.
\ee
 But now it is clear from \rf{eq:norm} that
 \be \pp{\eta} \int dx \, \epsilon \, \psi^2 = 0.
 \ee
 Therefore, we find that the variable $u(t)$ satisfies the equation of motion of a harmonic oscillator
 \be \ddot u = - \omega^2 u,
 \ee 
 where the frequency $\omega$ (could) be weakly time dependent. 
 
 Now, let us examine the energy of the wave equation. {We define it as}
 \be \label{eq:energy3} {\cal E} = \oh \int dx  \left[ \epsilon \left( \partial_t A \right)^2 +   \left( \partial_x A \right)^2 \right].
\ee 
 This can be found from the action $S$ in the usual way if $\epsilon$ is time independent. {It is also possible to show that the energy defined in this way is proportional to the actual textbook definition
 of the energy of the electromagnetic field, but we will not focus here on introducing the precise proportionality constant. In particular, the first term in the equation \rf{eq:energy3} represents the electric field energy, while the second term in the magnetic field energy.}

 Plugging in \rf{eq:eigen} into the energy, we find
 \be \label{eq:er}  {\cal E} = \oh \int dx  \left[ \epsilon \left( \dot u \psi + u \pbyp{\psi}{\eta} \dot \eta \right)^2 +   u^2 \left( \partial_x \psi \right)^2 \right] = \oh \left( \dot u^2 + \omega^2 u^2 \right) 
 +\int dx  \, \epsilon \dot u \psi u \pbyp{\psi}{\eta} \dot \eta.
\ee 
Here we used the relation, which follows from \rf{eq:eigen1}, 
\be   \int dx \left( \d_x \psi \right)^2 = \omega^2 \int dx \,  \epsilon \psi^2 = \omega^2. 
\ee
The last term in \rf{eq:er} can be rewritten in the following form
\be \frac{ \dot \eta}{2} \, \dd{t} \left( u^2 \int dx \, \epsilon \psi \pbyp{\psi}{\eta} \right) 
\ee
up to the terms quadratic in $\dot \eta$ which can be dropped. This term is not zero, but if averaged over one period of oscillations of $u$, assuming a very slow change in $\eta$, it becomes zero. 

Neglecting this term, the energy also reduces to the energy of the harmonic oscillator
 \be {\cal E} = \oh \left( \dot u^2 + \omega^2 u^2 \right). 
 \ee
 
 Now we recall that we are interested in the situation where $\omega$ slowly changes in time. Under these conditions, the adiabatic invariant remains approximately conserved. For a harmonic oscillator, it's equal to 
 \be I = \frac{{\cal E}}{\omega}.
 \ee
 We arrive at an interesting conclusion: the energy of the wave propagating in the presence of the slowly changing disorder changes
 proportionally to the possibly changing frequency of the wave. In turn, $I$ can be
 interpreted as the number of photons in 
 our electromagnetic wave. 

 Note that the analysis above assumes deep adiabatic evolution, where the adiabatic ansatz
 \rf{eq:eigen} is applicable. A more realistic 
 description of the solution takes into account
 that for any finite
 $\dot \eta$, $A$ becomes a linear 
 combination of the eigenfunctions $\psi$
 which correspond to frequencies within a window
 $\delta \omega \sim \dot \eta$.
 Taking this into account
 goes beyond the purely adiabatic theory developed 
 above, but is important in order to interpret
 the numerical simulations described below.

 To illustrate how a wave evolves under the adiabatically 
 increasing disorder we numerically solved the {discrete version of the } equation
 \rf{eq:wave}. We discretized space into 50000 points. We replaced the spatial derivatives in the equation \rf{eq:wave} by differences. For the function $V$, we generated an array of 50000 values, grouped into consecutive blocks of five identical entries (to simulate a small but nonzero disorder correlation length, effectively equal to 5 lattice spacings). Each block value was independently drawn from a uniform distribution
 ranging from $0$ to $1$. This
arrangement produces disordered $\epsilon$,
 with correlation length equal to $5$ lattice spacings. 

\begin{figure}[h!]
  \includegraphics[width=.99\linewidth]{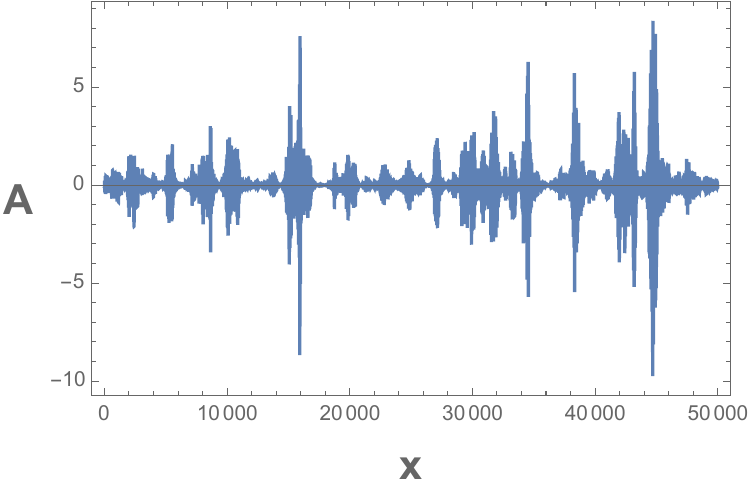}
\caption{
 Snapshot of the spatial dependence of the gauge potential after electric permittivity 
which varies randomly in space was adiabatically turned on. Normal  modes become localized. Note that the electric field is proportional to the time derivative of the gauge potential shown in this snapshot, while the magnetic field is proportional to the derivative with respect to the coordinate $x$. } 
\label{LocalizationAugust17}
\end{figure}

 {Thus the equation that we simulated took the following form
 \begin{equation} \label{eq:discreteequation}
 \partial_t \left[ \left( 1-\eta(t) V_x \right)
 \partial_t A_x \right]=
 A_{x+1}+A_{x-1}-2A_x
 \end{equation}
 where $x$ are now  integers labeling  lattice sites ranging from $1$ to $50000$ with 
 periodic boundary conditions.

We chose $\eta = t/50000$ for $0<t<50000$,
as well as $\eta=0$ for $t<0$,
to turn the disorder on adiabatically
starting at $t=0$.
Note that for $t<0$ this equation has exact
solutions in the form
\be \label{eq:inicondition} A_x = \cos\left(2 \pi k x/50000 - \omega_k t
\right),
\ee labeled by the integers $k$. Here
\begin{equation} 
\omega_k^2 = 2\left(1-  \cos(2 \pi k /50000)
\right).
\end{equation}

This solution represents a plane wave with
the wave vector $k$, which should be confined to take values $k \in \left( -25000, 25000\right]$ (in solid state physics terminology this corresponds to the lowest Brillouin zone). Now once the disorder is switched on, we anticipate that the higher
is the initial wave vector $k$, the stronger
the localization effects will be. 

To see strong localization effects, 
at $t=0$ we choose 
\be \label{eq:iniw} A_x = \cos \left(2 \pi k_0 x/50000 \right),\ee with $x$  varying from $1$ to $50000$, 
with the choice $k_0=1000$. This choice of $k_0$ ensures that
the initial wave length of the wave is short enough
for the wave to be localized at the end of the process and still long enough to contain $50$ lattice sites, which should be sufficient to approximate the behavior of the wave
equation continuous in space.  The initial condition for $\partial_t A$ at $t=0$ is chosen as
$\omega_{k_0} \sin\left(2 \pi k_0 x/50000\right),$
to conform to the equation \rf{eq:inicondition}
at $k=k_0$.

We solve the equation \rf{eq:discreteequation} numerically for $0<t<50000$ by implementing
the 4th order Runge-Kutta method with the
time step $\delta t= 0.01$. We plot the resulting function $A(x) \equiv A_x$ at the final
time $t=50000$, as shown in Fig.~\ref{LocalizationAugust17}. We see that the initial
plane wave \rf{eq:iniw} breaks up into ``lumps", just as we
anticipated. Now every solution of
the eigenstate equation \rf{eq:eigen1}
is localized nearby a randomly chosen site of the lattice. The wave, after 
disorder is adiabatically turned on, is a linear combination
of several of these eigenstates, the ones whose spectrum lies in the window $\delta \omega \sim \dot \eta$ around the main frequency
which adiabatically evolved from the
frequency of the initial wave $\omega_{k_0}$. Since
the spectrum of $\omega$ is closely spaced, even in the regime of $\dot \eta \ll 1$ quite a few of the eigenstates contribute to the
function $A(x)$. That's why it looks like a combination
of a number of the localized ``lumps". 

}

To summarize the results of this section, we considered a somewhat unusual, time-dependent setup for Anderson localization, when the wave is already inside the medium and random scatterers start growing within the medium.  We demonstrated that localization is indeed expected on scales  (\ref{calL}), and that during adiabatic switch-on, the number of photons within the system remains conserved.
In this heuristic approach, the growth of the amplitude of scatters is a prescribed function: this establishes the principle of localization inside the medium. 

As we demonstrate next in \S \ref{Clemmow1} using direct PIC simulation, the system in fact quickly evolves into a nonlinear stage, where one cannot easily separate the cause and effect in the plasma-EM wave interaction.

\section{PIC simulations of localization of EM wave already present inside plasma: the Clemmow frame}
\label{Clemmow1}

In many astrophysical applications, a nonlinear wave is generated by plasma itself, not impinging  from the  outside.  This  is very different from a conventional laser experiments.  The two set-up represent  somewhat different mathematical problems: boundary condition for external wave, and eigenvalue problem for a wave inside plasma \cite{taylor1965transmission}.

{In PIC simulations, it is typical not to initialize the setup with a self-consistent distribution of charges, currents, and \EM\ fields, especially of non-linear waves, as  fields and currents, evaluated at different spacial and temporal points,  should match at the start, {\bf but the exact nonlinear solution is often unknown}. There is actually a special case - that of superluminal EM waves  (\eg in unmagnetized plasma) where such a set-up is particularly simple, in what we call below {\it the Clemmow} frame \cite{1974JPlPh..12..297C}. }

\subsection{CP wave in Clemmow frame}
We start with the dispersion relation of a non-linear CP EM wave in pair plasma
\cite{1975OISNP...1.....A}
 \ba &&
  \omega ^2- k^2= \frac{2 \omega _p^2}{\gamma_\perp}
  \nn && 
\gamma_\perp = \sqrt{1+a_0^2}
  \label{disp11}
  \ea
  (Heuristically, it's a change of effective mass, $m_e \to  \gamma_\perp m_e$.)
To mind: $ \omega ^2- k^2$, $a_0$, $\gamma_\perp$, and  $\omega _p$  (as measured in the plasma rest-frame), are all Lorentz invariant quantities.

For super-luminal waves with refractive index $n< 1$,  one can make a boost with $\beta_x = k/\om = n$  to a frame where \Bf\  is zero, while \Ef\ is constant in space.
 In the boosted frame $K$ (quantities defined with subscript $0$), 
\ba && 
k = 0
\nn && 
\om  = \om/\gamma_x = \frac{\sqrt{2} \omega _p}{(1+a_0^2)^{1/4}}
\nn && 
|E| = a_0 \om 
\nn && 
\gamma_x = \frac{1}{\sqrt{1-n^2}} = \frac{  (1+a_0^2) ^{1/4}} {\sqrt{2} } \frac{ \om }{\om_p}
\nn &&
\beta_x =  n= \left(1+ 2 \frac{ \om_p^2}{ \sqrt{1+a_0^2} (  c k_x)^2} \right)^{-1/2} 
\nn &&
\gamma_{\rm tot} = \gamma_x  \gamma_\perp = \frac{  (1+a_0^2) ^{3/4}} {\sqrt{2} } \frac{ \om }{\om_p}
\ea
Parameter $\gamma_{\rm tot}$ is the total \Lf\ in the $k$ frame. 

In $K$ the current is compensated by the changing \Ef, 
\be
{4 \pi} {\bf j}+ \partial_t \E =0
 \ee
 (using $c=1$). In unmagnetized pair plasma the current 
 is  equally  contributed  by electrons and positrons 
\be
{\bf v_{\perp, p}} = - {\bf v_{\perp, e} } = \frac{ {\bf j}}{2 n_0}
\ee

 Importantly, for 1D motion the parallel \Lf\ $\gamma_x$ factors out - we can write  in $k$:
 \ba &&
 \om ^2 = \om^2-k^2 = 2 \frac{\gamma_x \om_p^2 } {\gamma_x \gamma_\perp } =  2 \frac{ \om_{p,0}^2 }{\gamma_x \gamma_\perp } \to   2 \frac{ \om_{p,0} ^{\prime,2}} { \gamma_\perp }
 \nn &&
 \om_{p,0} ^{\prime,2}= \frac{ \om_{p,0}^2 }{\gamma_x}
 \ea
 This is just a re-definition of plasma frequency (seen either as a re-definition of plasma density or ``effective'' mass). 

 Thus, neglecting parallel motion, a  nonlinear CP wave {\it inside} pair plasma is given by
 \ba &&
\v_p = \{-\cos (\omega_0  t),-\sin (\omega_0  t),0\} \frac{a_0}{\sqrt{1+a_0^2}}
\nn &&
\v_e = - \v_p
\nn &&
A_0= \{\cos (\omega_0  t),\sin (\omega_0  t),0\} a_0
\nn &&
\E_0  = \{\sin (\omega_0  t),-\cos (\omega_0  t),0\} a_0 \omega
\nn && 
\om  = \frac{\sqrt{2}  }{(1+a_0^2)^{1/4}} \om_{p,0}
\label{1}
\ea
Below in this section we drop the subscript $0$.

Effectively,  transformation to the Clemmow frame gets rid of the fast oscillating and propagating wave phase; besides spatially homogeneous oscillations, any particle dynamics is a ``slow'' (\eg modulational) dynamics in the lab frame. 

The Clemmow frame is not a Lorentz-boosted frame where the beam is at rest. Plasma may still drift along the wave's propagation, but that drift can be parametrized out (in 1D) in a new definition of plasma density. The Clemmow frame is also not the group velocity frame.

\subsection{Set-up of PIC simulations in the Clemmow frame}

We performed  PIC simulation of the EM wave in the Clemmow frame. Numerical set-up is somewhat unusual: the laser-oriented code EPOCH  uses an explicit expression for the wavelength of the driver, while in the Clemmow frame $\lambda \to \infty$. The important spatial parameter is the skin depth $\delta = c/\om_p$. To convert, 
\be
\lambda  \equiv   2 \pi  \sqrt{\frac{n_{cr}}{n}} \delta \to   2 \pi \delta
\label{lambdatodelta}
\ee
(factor $\sqrt{2}$ not included; in pair plasma the physical skin depth is $\delta ^{(\pm)} = \delta /\sqrt{2}$.) 

In the code, we set $n=n_{\rm cr}$. We stress that $n=n_{\rm cr}$ does not mean reflection: the wave is {\it already} inside plasma;  in the Clemmow frame, intensity and density determine frequency.   At time $t=0$ there is  spatially   constant \Ef (along $y$) and  spatially constant $v_\pm$ 
(along $z$ for CP and  along $y$ for LP). Initial velocities (in the CP case) are tuned to the corresponding analytical solutions.  External laser intensity is set to zero.

{Importantly, since \Bf\ is vanishing, the transformation to the Clemmow frame avoids the numerical problem that electric and magnetic fields in PIC codes are evaluated at different spatial grids. Temporal mismatch between currents and fields can be made small enough by choosing sufficiently high $n_x$.
}

\subsection{PIC simulations of self-localization of EM wave in the Clemmow frame}

\begin{figure}[h!]
  \includegraphics[width=.49\linewidth]{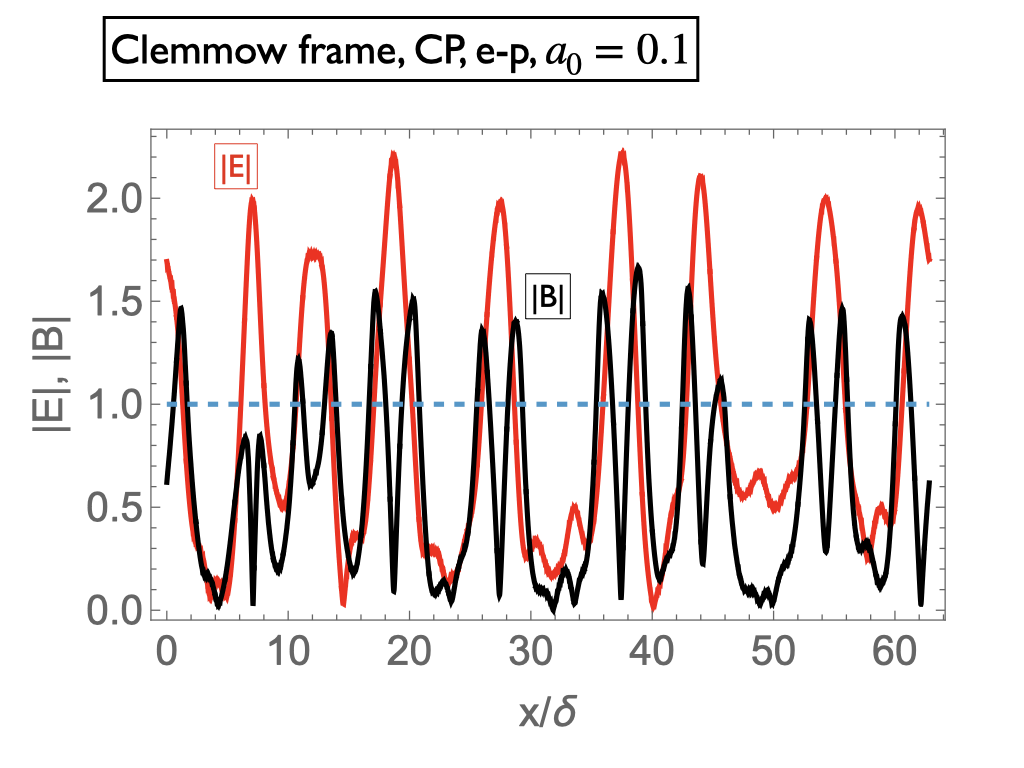}
   \includegraphics[width=.49\linewidth]{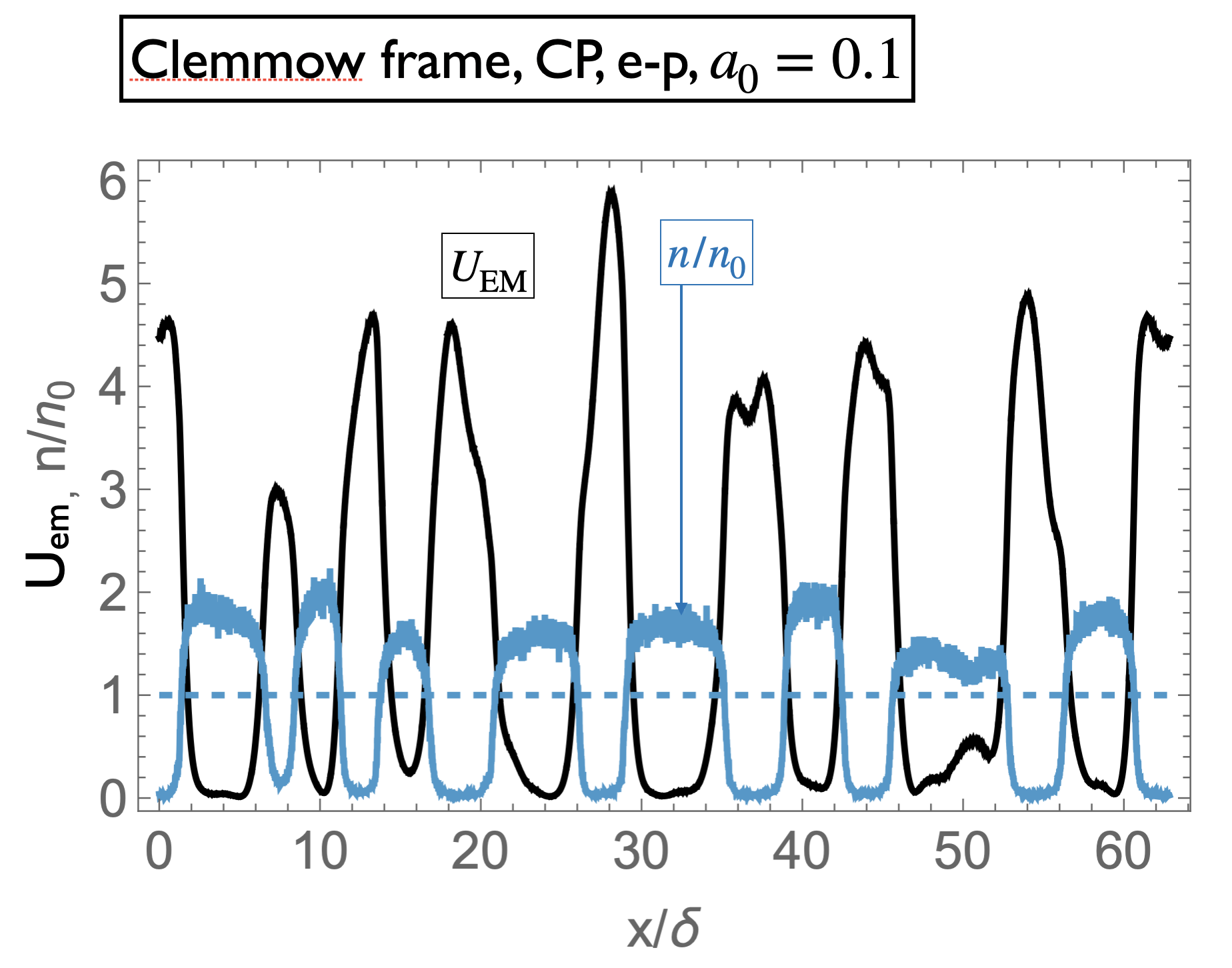}
\caption{PIC simulations of Anderson localization of light in pair plasma in  Clemmow frame, CP $a_0=0.1$, $e^\pm$ plasma.
Left panel: values of electromagnetic fields  $|E|$ (red)  and $B$ (black). (Parallel component for \Ef\ is small, and is not included.) 
Right panel: total \EM\ energy density $U_{\rm EM}$ (black) and plasma density (dark blue). One clearly sees formation of localized structures on scales $\sim $ few skin depth, where high EM intensity is associated with low plasma density.  } 
\label{Clem01-1}
\end{figure}

Initially, \Ef\ and density are spatially constant, while \Bf\ is zero. Quickly, the system develops regions of high \EM\ field intensity associated with low densities, and intermitted with high density ``walls". 
The most exemplary case is $a_0=0.1$, Figs. \ref{Clem01-1}. This is   Anderson's self-localization of EM waves already present inside  pair plasma, reproduced via PIC simulations.

\begin{figure}[h!]
  \includegraphics[width=.99\linewidth]{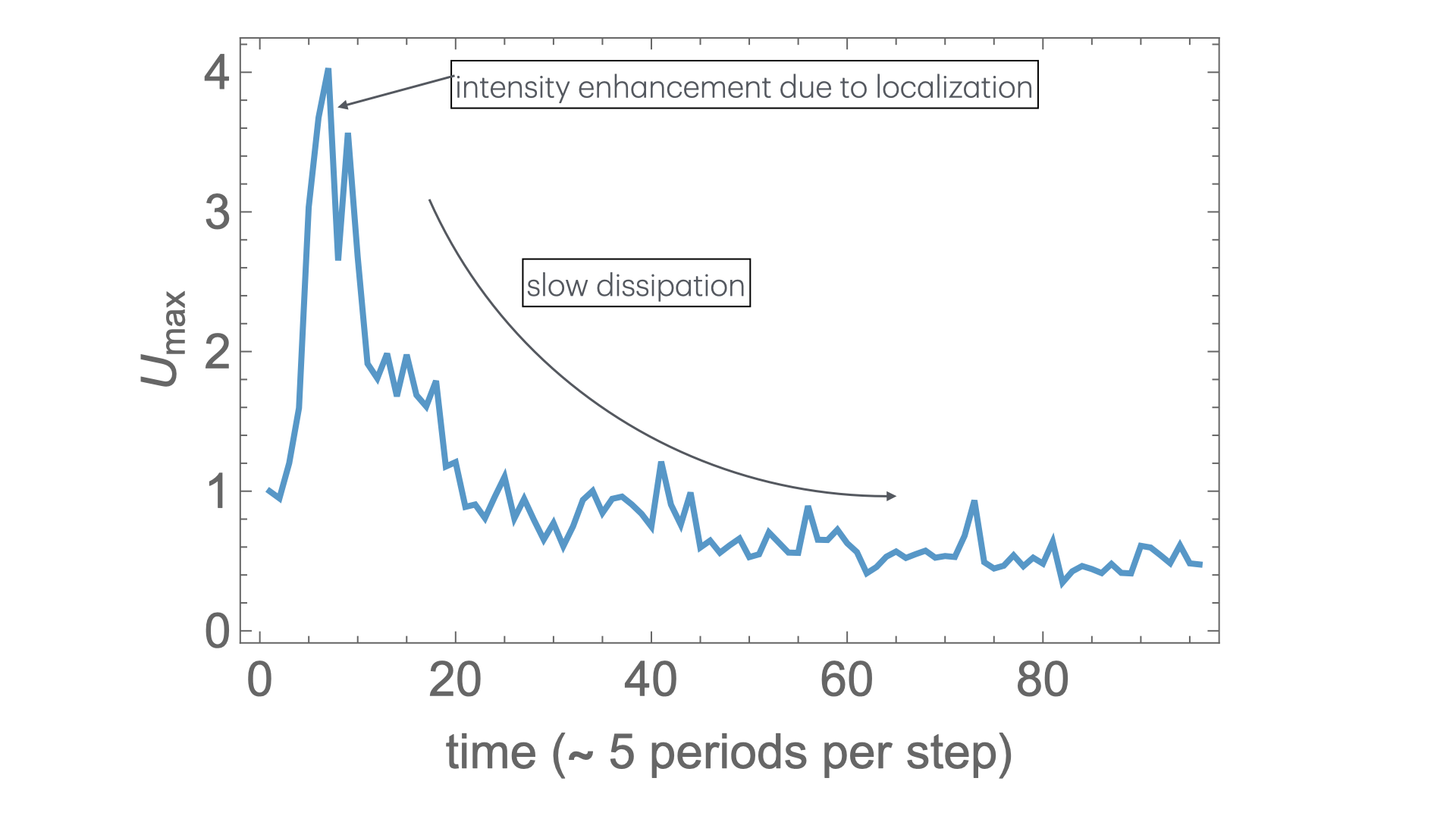}
\caption{Example of enhancement of local wave energy density due to effects of localization, Clemmow frame, $a_0=1$. Plotted is a maximal value of energy density in the 1D domain of 100 effective wavelengths.} 
\label{enhance}
\end{figure}

Wave localization leads to local enhancement of intensity. In  Fig. \ref{enhance}, we demonstrate an enhancement of local intensity due to waves' localization. In the Clemmow frame, $a_0=1$, we plot the maximal value of the \Ef\ at each point as a function of time, normalized to the initial field. We observe an initial spike due to localization, followed by a slow dissipation. This is the only run - no special efforts were made to increase the peak intensity. We hypothesize that occasionally, much larger intensity enhancements are possible.

At large density/EM perturbations, we are likely to switch to ponderomotive localization, where many random-phased waves provide a grad $E^2$ force that pushes plasma out.

\begin{figure}[h!]
  \includegraphics[width=.99\linewidth]{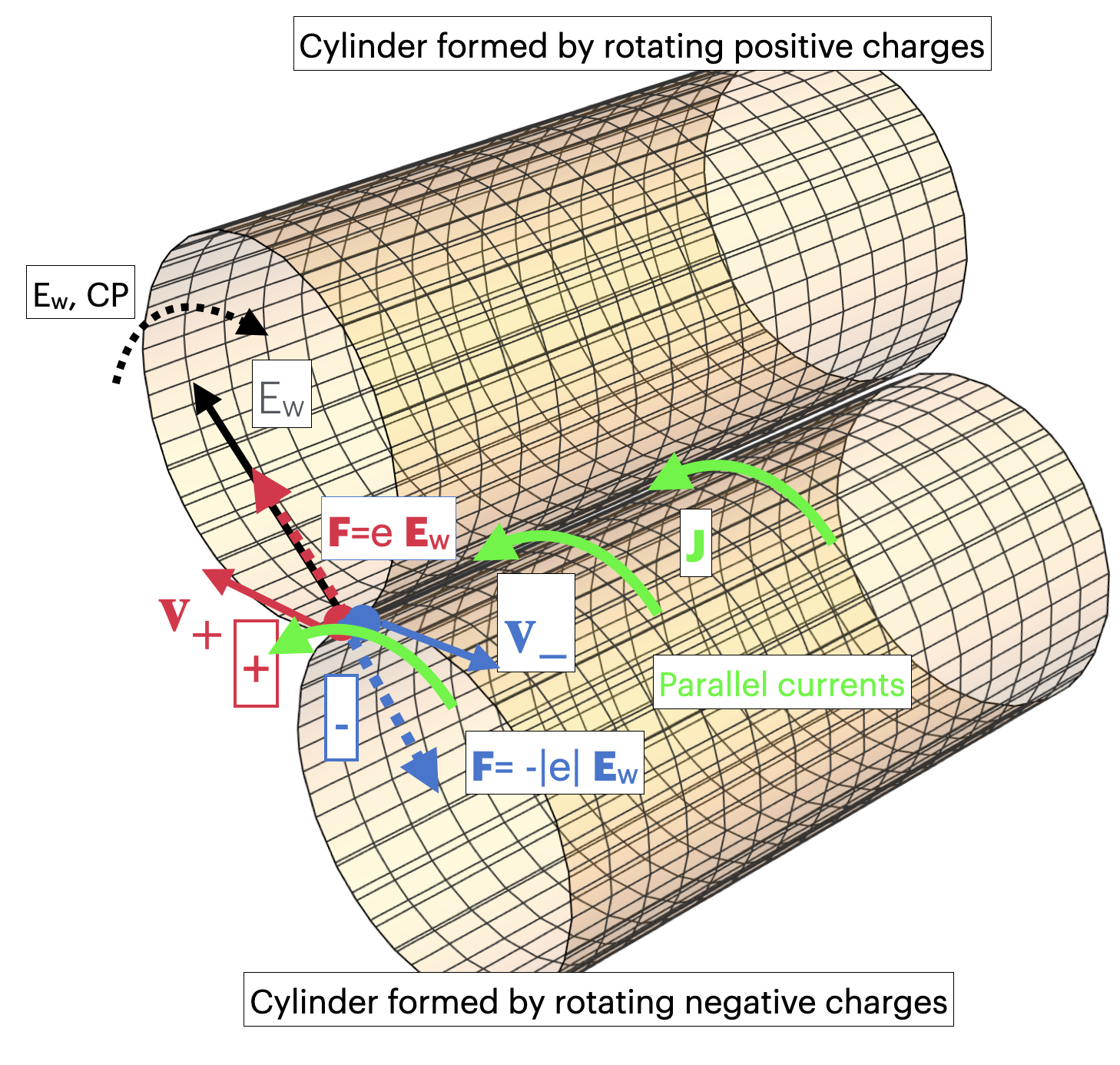}
\caption{Illustration of particle trajectories and currents for CP  wave in the Clemmow frame. The \Ef\ is always perpendicular to the particle's velocities, while at each point $v_\pm$ are counter aligned. Particles of opposite sign  are counter-streaming at each point. moving along two cylindrical surfaces.   The resulting modulational instability of waves inside plasma belongs to a general class of  Weibel-like instabilities, ``parallel currents attract". Current bunching will occur in the direction along the cylinders. } 
\label{weibel}
\end{figure}

In the Clemmow frame,  for CP  wave, the setup  and the resulting dynamics resemble Weibel instability,  $\sim$ parallel currents attract, Fig. \ref{weibel}.

  \subsection{Polarization structure of localized EM waves}

\begin{figure}[h!]
   \includegraphics[width=.99\linewidth]{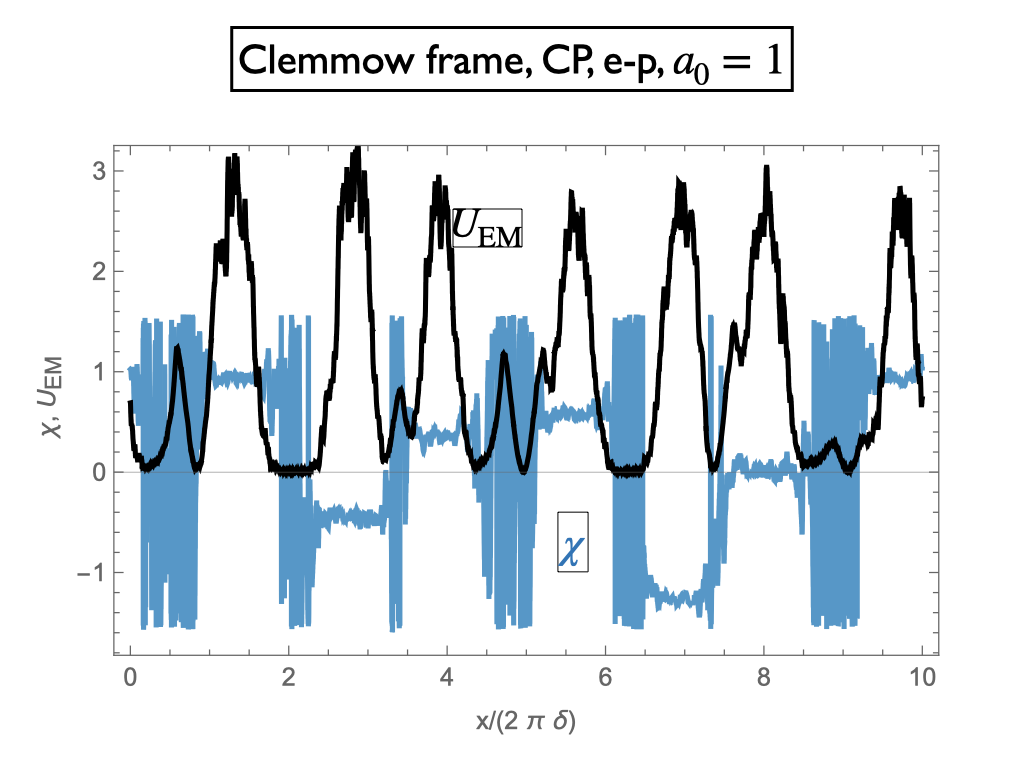}
\caption{Comparison of profiles of polarization angle $\chi$  (dark blue curves) and EM  energy density  (black curves).} 
\label{Pi}
\end{figure}

The polarization structure of localized EM waves is quite unusual. In Fig. \ref{Pi} we compare profiles of EM  energy density  $U_{\rm EM}$ and  position angle (PA) calculated as
\be
\chi =\arctan (E_y/E_z)
\ee

  Within the bright region, the position angle PA remains approximately constant! Roughly, this corresponds to a linearly polarized wave. {\it Thus, a circularly polarized  driver produced linearly polarized  localized waves}. The value of PA varies randomly between the bright regions. 

\section{Discussion}

In this work, we discuss  Anderson self-localization of \EM\ waves in pair plasma. The most surprising  example is a reflection of a mildly nonlinear  EM pulse by an underdense pair plasma. This is achieved through creation of irregular fluctuations of density by the leading edge of the pulse, and corresponding fluctuations of the dielectric constant. By a classical result of \cite{doi:10.1137/1104038}, such media reflect EM waves. An important new ingredient here is that the density/dielectric permittivity fluctuations are created by the wave itself.

Our work extends the results of  \citep{2025arXiv250906230T}, which concentrated on $a_0 \geq 1$  regime, to mildly non-linear regime $a_0 \leq 1$. The high $a_0$ regime clearly shows the effect of reflection on EM waves by under-dense plasma. In this regime, the system quickly goes highly nonlinear, and  plasma accelerated to relativistic \Lfs. In the mild regime $a_0 \leq 1 $ we can trace more clearly the underlying cause: parametric generation of  large density spikes, and ensuing effects of Anderson localization.

While the ``nose'' part of the pulse propagates as expected from the linear theory, the bulk of the pulse is reflected (for limited pulse duration) or is localized near the boundary (for long pulse duration). In the latter case, there is a dark region between the surface-localized EM structures, and a propagating  ``nose''.  

The effect we observe is intrinsically kinetic.  Like in the SASE regime of FEL \citep{freund1986principles}, the  seeding  comes from the discrete nature
of electrons in the beam. Discreteness  here means statistical fluctuations in the electron density.  For example, in the case of SASE FEL, in the fully continuous limit, the electron beam is just a constant current that does not radiate. Also, in the fluid limit back-scattering is completely absent, due to destructive interference.

One  also expects the formation (only if temporarily) of the transparent band gaps, possibly exhibiting  narrow-band amplification \citep[\eg][]{2016PhRvL.116v5002L}.

A new effect, if compared with reflection by a varying dielectric, appears in pair plasma. 
In addition to random fluctuation of the value of the dielectric constant, locally, within the density sheets, the plasma may become over-dense. As a result, a  wave becomes evanescent, contributing to the effect of localization.

{
An important difference from the conventional time-invariant Anderson localization is that fluctuations of the medium induced by the wave are both spatially and {\bf temporarily}  random. Conventional wisdom   \citep{1958PhRv..109.1492A,2014PhRvL.112b3905S} says that temporal variations in the potential
reduce the coherence of the scattering process, and thus tends to diminish localization effects. 

In our  case, the "scatterers" are not external independent random objects, but self-induced structures that are phase-locked to the field. That is, there is a time dependence here, but it is correlated/phase-locked, and therefore it is not obliged to kill the interference in the same way as "white" temporal noise. (Fig.\ref{dens0019_fft_zoom} illustrates the presence of coherent structures in the seemingly random density field.
}

A somewhat different set-up involves a wave already inside a medium - when slow variations of parameters (\eg changing guide \Bf\ and/or density) bring the system into a regime of self-localization. In this case, a wave separates into bright and dark regions. The initial momentum of the EM wave is transferred to the medium.
As the system is highly nonlinear, at a developed stage of localization, one cannot distinguish Anderson localization (in the sense of importance of waves' phases) from ponderomotive localization (when only the averaged amplitude is important).  

Previously, doing PIC simulations,  in Ref. \cite{Huang_2021} relativistic-induced opacity of electron–positron plasmas was  noticed.   Our more extensive results seem to be consistent with  simulation of \cite{Huang_2021}, but not with  their interpretation, (that reflection is due to density pile-up): density pile-up, exceeding the expected, is the result of  wave the reflection, not the cause.

 Here we limit ourselves to a 1D system - physically a radial distance from a source. This approach may be justified by noting that corresponding intrinsic effects are related to plasma kinetic scales, like skin depth, while transverse extensions are likely to be macroscopic. Of course, 1D is an approximation, as relativistic effects in expanding pulsar wind greatly increase relevant physical scales in the observer frame.   Wave localization in transverse direction  \citep[\eg][]{1989PhRvL..62...47D,2013NaPho...7..197S} may lead to formation of collimated radiation beam (instead of conically expanding). Possible application of transverse localization to pair plasma requires independent  study.

{

Next, we briefly comment on the possible importance of the results for applications to the physics of astrophysical FRBs, leaving a more targeted discussion for a follow-up paper.  The processes discussed here are likely to be important in \mss\ and relativistic winds of \NSs.  (Effects of external  \Bf\ on the wave localization will need to  be considered.)

Self-localization in the wind is likely to be especially important. First, in relativistic pulsar wind moving with \Lf\ $\gamma_w$, the localization in the lab frame can take a long time, as long as  $\propto \gamma_w^2 $ times longer than in the wind frame. This potentially can erase the intrinsic periodicity related to the spin of the host \NS. 

Second,  EM waves trapped/localized within the wind are still moving relativistically in the lab frame; they eventually may convert into escaping pulses of radiation. 

Thus, one may envision a two-stage process: generation of mildly intense, non-localized EM within the \ms, and further modification due to propagation/localization. 

Observationally, several points are worth highlighting now:
\begin{itemize}
    \item Wave localization introduces extra temporal randomness (in addition to the operation of the primary source of EM waves). This may explain a clear lack of periodicity in FRBs (as expected from a rotating \NS).
     \item Strength of localization decreases both at low and high frequencies - FRBs are indeed seen in a limited range of frequencies.
   \item Variable Polarization structure  \citep[polarization properties of FRBs defy simple classification][]{2019MNRAS.487.1191C,2019A&ARv..27....4P}.  As we demonstrate, even a CP driver in a pair plasma can produce a linearly polarized localized EM structure. 
\end{itemize}
}

ML is mostly grateful to  Alexey Arefiev and Kavin Tangtartharakul for numerous discussion and code advise. Special acknowledgment is due to Alexey Mohov. 
 This research was supported in part by grant NSF PHY-2309135 to the Kavli Institute for Theoretical Physics (KITP). This research was also  supported by the Simons Collaboration
on Ultra-Quantum Matter, which is a grant from the Simons Foundation (651440, VG).
We would like to thank participants at the program "Frontiers of Relativistic Plasma Physics" for numerous discussions. Special acknowledgments are due to Paulo Alves, Frederico Fiuza, Thomas Grismayer,  Mikhail Medvedev,  Alexander Philippov, Robert Russel,  Luis Silva, Lorenzo Sironi, Navin Shridhar,  Anatoly Spitkovsky,  Arno Vanthieghem, Marija Vranic and  Dmitri Uzdensky. ML aslo would like to thank   Natalia Berloff, Amitava Bhattacharjee,  Sergey Bulanov, Yong Cheng and  Yuli Lyanda-Geller.

 \bibliographystyle{apj} 
\bibliography{./BibTex.bib,./BibTexShort.bib,./referencesVictor.bib} 
 
\appendix

\section{Simulations' details and convergence}
\label{Epoch}

The simulations were performed using the EPOCH code \cite{Arber:2015hc}, running on the on the Purdue cluster {\it Negishi}. Unless stated otherwise,  parameters were $n= 0.0125 n_{\rm crti}$, {typical values of particles per cell are $n_p=100$, cells per wavelength are $n_x=100$, tough longer convergence runs use smaller $n_p=n_x=10$; results are generally consistent. Plasma occupies region $x> 0$, the EM source is located at $x= - 50 \lambda$. Left boundary condition is ``simple laser'', right boundary condition is ``open''. We keep track that the laser does not reach the right boundary. (Exception is simulations presented in Fig. \ref{PIC-plus-minus}, where both  boundaries are ``simple laser'';  in this case, for outgoing waves the boundary conditions are equivalent to ``open'').}
In all simulations presented here, in the case of e-i plasma,  ions were taken to be immobile.

Few comments on numerical demands are due:
\begin{itemize}
\item  Anderson localization does require keeping the phase of the waves correctly, which is challenging numerically. 
\item  One does expect large fluctuations of intensity in the localization regime - this complicates the measurements of the localization length.
\end{itemize}

The crucial issue for convergence tests is that the effect of Anderson self-localization depends on the fluctuations of the density. Hence, we do not expect convergence for arbitrary different simulation parameters.
Instead, we expect that the onset of localization depends on the total number of particles as square root. 

\begin{figure}[h!]
\includegraphics[width=0.99\linewidth]{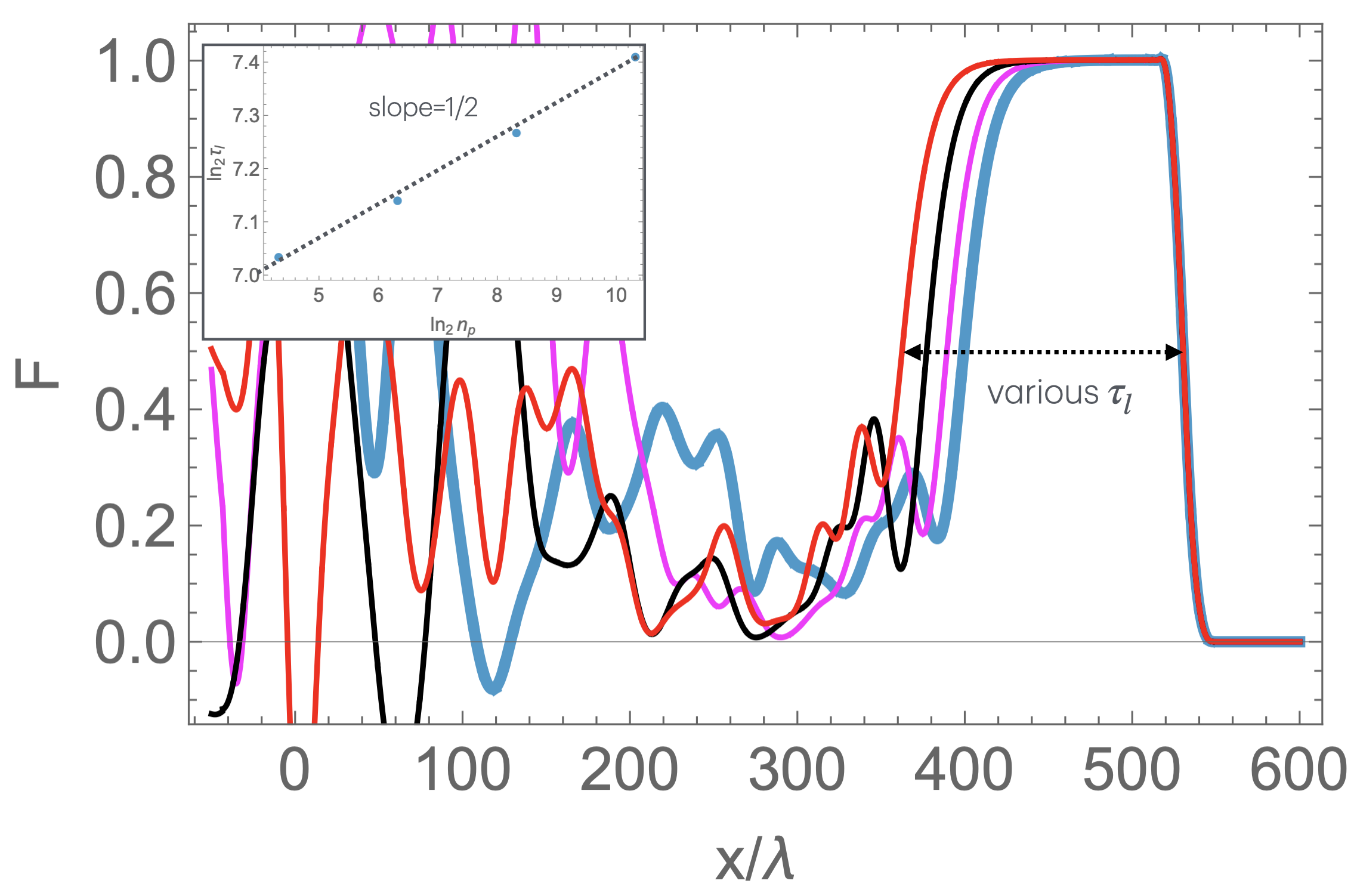}
\caption{Poynting  flux for fixed $n_x=30$ and variable $n_p$. The insert shows dependence of localization time $\tau_l$ (measured as the duration of the unperturbed pulse), on number particles per cell $n_p$; it is highly consistent with expected   $\tau_l \propto \sqrt{n_p}$.}
\label{testnp}
\end {figure}

The only stable feature, present in all simulations, is ``the nose'': a limited duration part  of the pulse that does propagate into pair plasma.
In Fig. \ref{testnp} we plot the duration of this feature as a function of  the parameter $n_p$, {\bf  number particles per cell}.  We indeed observe the
expected $\sqrt{n_x} $ dependence (see insert).  We also verified that: (i) fixing $n_x $ and varying $n_p$ produces similar scaling; (ii) runs with the same total number of particles $n_x n_p$ produce similar profiles.   We conclude that our simulation results are consistent with the expectation.

Most simulations in this work are done for cold plasma,  the code   under-resolves the Debye radius. (In \S \ref{Tstabilization} we explore the effects of initial temperature and  we  find that  small initial temperature, $\Theta \ll \Theta_{a_0}$
does not affect the results of simulations, Fig. \ref{T0}.)
We have verified that numerical heating is minimal, and that resolving the Debye scale does not change the final results.

\section{Reflection of EM waves by randomly-located density  delta-layers}
\label{deltaspike}

{As a complementary look at the effect of Anderson self-localization of light, we  consider a complementary problem, approximating density fluctuation as
sheets with a variable dielectric tensor. This mimics logarithmically divergent charge-neutral density peaks  expected due to the beat of the driver with the back-scattered wave in pair plasma  \citep{2021ApJ...922..166L}. Importantly, here  we assume stationary scatterers.}

Let's first look at the simplest dielectric response, approximating density fluctuation as
sheets with a variable dielectric tensor. 
Each density sheet can then be parametrized by the integrated density
\ba &&
\sigma_n  = \int n (x) dx, [ {\rm cm^{-2}}]
\nn &&
n (x) = \sigma_n \delta(x) , [ {\rm cm^{-3}}]
\ea
(in parenthesis, we give the corresponding dimensions).

For   a given static density distribution, the plasma response tensor $\sigma _{ij} =\sigma \delta _{ij} $ (not to be confused with surface density $\sigma_n$) is 
\ba &&
j_i = \sigma _{ij}  E_j = \sigma \delta _{ij} E_j
\nn && 
\sigma =   i \frac{ e^2  n}{\om m_e}  =  i \frac{ e^2   \sigma_n }{\om m_e} \delta(x) 
\nn &&
\epsilon= 1-\ell  \times  \delta(x)
\nn &&
\ell  = \frac{4 \pi e^2}{m_e \om^2} \sigma_n,  [ {\rm cm}]
\label{sigma}
\ea

Importantly, the form of $\epsilon$ implies that the potential associated with each density sheet is  attractive. For an attractive delta-potential, even a single layer has a bound state. (Another related example is a quantum mechanical Dirac comb with a missing pin  -  it also has  a localized wave function.) 

The quantity $\ell $ in (\ref{sigma}) has a dimension of length. Since the only other dimensional quantity is the wavelength of the incoming wave, one can rescale 
\be
\ell = 2 \frac{c}{\om} \Gamma
\ee
where $ \Gamma$ is the dimensionless power of the scattering screen.  (Factor of $2$ makes the corresponding relations for the scattering matrix  (\ref{S}) particularly simple.)

For a wave falling onto a static dielectric  $\delta$-layer, the wave equation is
\ba &&
\partial_x^2 E = \partial_t^2 D 
\nn &&
D = \epsilon E 
\ea
with $\epsilon$ given by (\ref{sigma}).

Let a wave fall on the layer from  $x< 0 $:
\ba && 
E=  e^{- i (\om t - kx ) } +  r  e^{- i (\om t +  kx ) }, x< 0
\nn &&
E=  t e^{- i (\om t - kx ) }, x> 0 
\ea
where $r$ and $t$ are complex reflection and transmission coefficients for amplitudes.

Conditions
\ba &&
[E] =0
\nn && 
[E']=  \om^2    \ell   E_{x=0}
\ea
give complex  transmission and reflection coefficients
   \ba &&
{   t=\frac{i}{i-\Gamma}}
   \nn &&
  {r= \frac{\Gamma}{i-\Gamma}}
   \label{rt}
   \ea
{(we use here  standard notation of small-$t$ for transmission coefficient; it should not be confused with time).}

One can then define the scattering matrix
   \be
   S =
   \left(
\begin{array}{cc}
 \frac{i}{i-\Gamma} & \frac{\Gamma}{i -\Gamma} \\
 \frac{\Gamma}{i -\Gamma} & \frac{i}{i-\Gamma} \\
\end{array}
\right) 
\label{S}
\ee
It is unitary, $S^\dagger S = I$.
The scattering matrix (\ref{S})   acts on a vector
\be
  \left(
\begin{array}{c}
E_+ \\
E_-
\end{array}
\right)
\ee
where  $E_\pm $ are the amplitudes of right/left propagating modes.

In passing, we note that for two screens separated by $x_0$, zero reflection corresponds to 
   \be
   \cos \left(2 k x_0\right)= - \frac{1- \Gamma^2}{1+ \Gamma^2}.
   \ee
         In the case of weak scattering   $\Gamma \to 0 $ we then recover the condition for {quarter-wavelength anti-reflection coating}
$   
 x_0={\lambda }/{4}
$.

As an illustrative results, 
given the scattering matrix (\ref{S}) we can construct a model of a  {\it static}  1D plasma layer with randomly distributed scatterers.  At each layer, a wave gets reflected/transmitted with a complex scattering matrix. (Static is an important approximation here.) On the left boundary, there is only a right-propagating infalling wave; at the right boundary, there is only a transferred, right-propagating wave. In each interval between $n$th and $ n+1$th wall, we can then solve for the amplitudes of right/left propagating waves $E_\pm$. Numerically, a solution is possible for a number of scatterers $ N\sim$15.

\begin{figure}[h!]
 \includegraphics[width=.99\linewidth]{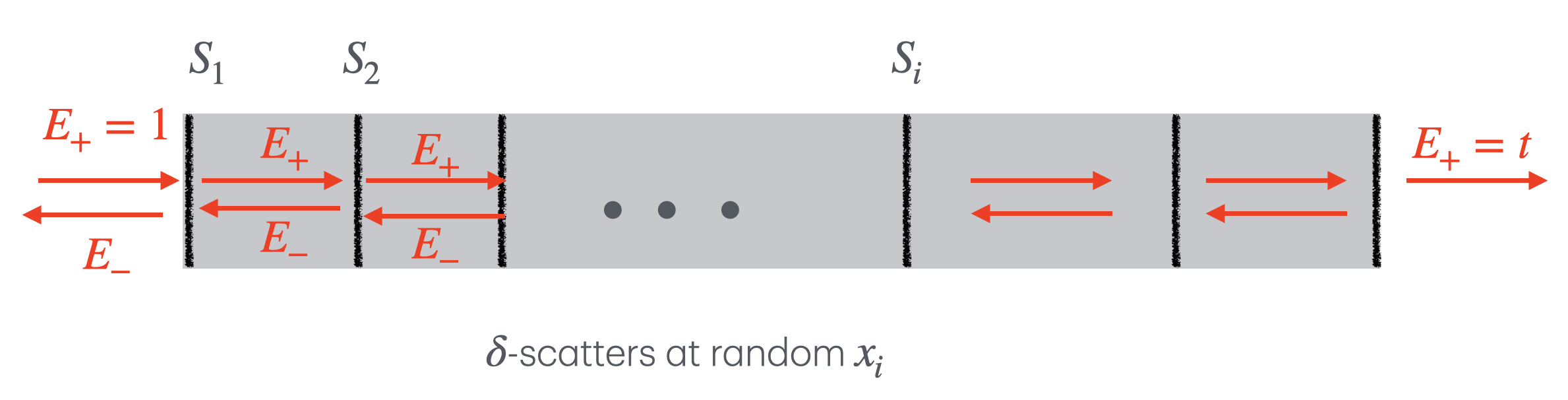}
  \includegraphics[width=.99\linewidth]{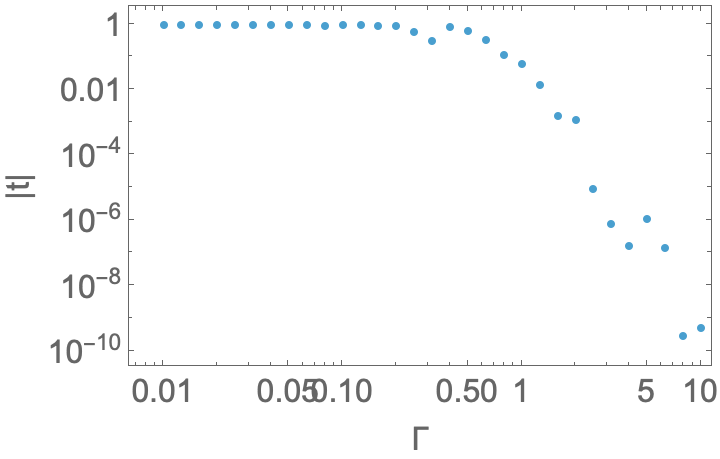}
\caption{Top:  1D model of random scatterers with scattering matrix (\ref{S}). Bottom:  resulting  transmission coefficient $|t|$ (for amplitude, not intensity) as a function of dimensionless strength $\Gamma$. The total length of the layer is 10 wavelengths, 12 scatterers total.} 
\label{toflambda0}
\end{figure}

In Fig. \ref{toflambda0}, we plot transmission probability as a function of $\Gamma$. As the intensity of each scatter approaches unity, the wave is nearly completely reflected. In essence, this is just an example of Anderson localization by a collection of static 1D scatterers.

In Fig. \ref{toflambda0} no averaging over random distributions was made (each run has a single random distribution of scatters). This explains the fluctuations in the transmission coefficient $t$. We also note that for $t\to 0$, fluctuations of $t$ become large. This is an expected feature of Anderson localization of light \citep{1992SvPhU..35..231K}

Our result is a straightforward application of Anderson localization, diffusion with a phase. We just  demonstrated it with a particular properties of scattering matrix (\ref{S}).

\end{document}